\documentclass[12pt,a4paper]{article}
\usepackage{mathtools}
\usepackage{mathrsfs}
\usepackage{amsmath}
\usepackage{tikz-feynman}
\usepackage[utf8x]{inputenc}
\usepackage{amssymb}
\usepackage{slashed}
\usepackage{bbm}
\usepackage{bm}
\usepackage{color}
\usepackage[a4paper,top=3cm,bottom=3cm,left=2cm,right=3cm,bindingoffset=5mm]{geometry}
\usepackage{booktabs} 
\usepackage{tikz}
\usepackage{cancel}
\usepackage[bottom]{footmisc}
\usepackage{float}
\usepackage{todonotes}
\usepackage{jheppub}
\usepackage{bbold}
\usepackage{comment}

\usetikzlibrary{shapes.misc}
\tikzset{
    point/.style={
    draw=black,
    cross out,
    inner sep=0pt,
    minimum width=4pt,
    minimum height=4pt,
    },
}

\usepackage{hyperref}

\hypersetup{
colorlinks=true,         
linkcolor=blue,          
citecolor=red,        
urlcolor=cyan            
}

\renewcommand{\[}{\begin{equation}\begin{aligned}}
\renewcommand{\]}{\end{aligned}\end{equation}}

\DeclareMathOperator{\Tr}{Tr} 

\newcommand{\wb}{{\bar{w}}}
\newcommand{\Pd}{\buildrel{\leftrightarrow} \over {P}}

\newcommand{\alphadot}{\dot{\alpha}}
\newcommand{\betadot}{\dot{\beta}}

\newcommand{\lambdat}{\tilde{\lambda}}

\title{On the anomaly interpretation of amplitudes in self-dual Yang-Mills and gravity}
\author[a]{George Doran,}
\author[a]{Ricardo Monteiro,}
\author[b]{Sam Wikeley}

\affiliation[a]{Centre for Theoretical Physics, Department of Physics and Astronomy, \\ Queen Mary University of London, E1 4NS, United Kingdom}
\affiliation[b]{Department of Physics and Astronomy, Uppsala University, \\ Box 516, 75120 Uppsala, Sweden}

\emailAdd{g.e.b.doran@qmul.ac.uk}
\emailAdd{ricardo.monteiro@qmul.ac.uk}
\emailAdd{sam.wikeley@physics.uu.se}

\abstract{
We investigate the integrability anomalies arising in the self-dual sectors of gravity and Yang-Mills theory, focusing on their connection to both the chiral anomaly and the trace anomaly. The anomalies in the self-dual sectors generate the one-loop all-plus amplitudes of gravitons and gluons, and have recently been studied via twistor constructions. On the one hand, we show how they can be interpreted as an anomaly of the chiral U(1) electric-magnetic-type duality in the self-dual sectors. We also note the similarity, for the usual fermionic chiral anomaly, between the 4D setting of self-dual Yang-Mills and the 2D setting of the Schwinger model. On the other hand, the anomalies in the self-dual theories also resemble the trace anomaly, sharing the same type of non-local effective action. We highlight the role of a Weyl-covariant fourth-order differential operator familiar from the trace anomaly literature, which (i) explains the conformal properties of the one-loop amplitudes, and (ii) indicates how this story may be extended to non-trivial spacetime backgrounds, e.g.~with a cosmological constant.
Moving beyond the self-dual sectors, and focusing on the gravity case, we comment on an intriguing connection to the two-loop ultraviolet divergence of pure gravity, whereby cancelling the anomaly at one-loop eliminates the two-loop divergence for the simplest helicity amplitudes.

}

\begin{document}

\begin{flushright}
QMUL-PH-23-26 \\
UUITP–38/23
\end{flushright}

\maketitle


\section{Introduction}
\label{sec:intro}

The main motivation for this paper is to explore the similarities between the integrability anomalies of the self-dual sectors of Yang-Mills and gravity, and the more familiar stories of the chiral and trace anomalies, particularly regarding non-local quantum corrections to the action.

The self-dual sectors are well known to have simple scattering amplitudes, non-vanishing only at one-loop for all-plus external helicities $(++\cdots+)$.\footnote{In our conventions, particles in a scattering amplitude are incoming (so that momentum conservation reads $\sum_i p_i=0$), and a positive-helicity polarisation is associated to a self-dual field. We note that the 3-point tree amplitudes with helicities $(++-)$ also arise, and play an important role if one relaxes the restriction of real kinematics in Lorentzian signature.} Behind the vanishing of the tree-level amplitudes, lies the integrability of the classical self-dual theories. The one-loop all-plus amplitudes are also special: they are rational functions with simple soft and collinear behaviour, in both Yang-Mills and gravity. These properties provided the guidance for constructing explicit $n$-point formulas for the amplitudes \cite{Bern:1993qk,Mahlon:1993si,Bern:1996ja,Bern:1998xc,Bern:1998sv}. The formulas are strongly suggestive of the existence of a simple one-loop quantum correction to the action, incorporating the loop-level effects. The idea that the one-loop amplitudes should be interpreted as signalling an anomaly of the classical integrability goes back to \cite{Bardeen:1995gk}; see also \cite{Chalmers:1996rq,Cangemi:1996rx}. Indeed, the amplitudes arise from an $\epsilon/\epsilon$ contribution in dimensional regularisation. This idea has been the subject of several investigations over the years, e.g.~\cite{Brandhuber:2006bf,Brandhuber:2007vm,Boels:2008ef,Krasnov:2016emc,Nandan:2018ody,Chattopadhyay:2020oxe,Chattopadhyay:2021udc}. It has recently seen a beautiful realisation in the twistor-space constructions of \cite{Costello:2021bah,Bittleston:2022nfr}; see also \cite{Costello:2022wso,Costello:2022upu,Bu:2022dis,Bittleston:2022jeq,Bu:2023vjt}. The goal of \cite{Costello:2021bah,Bittleston:2022nfr} was to obtain quantum-integrable `improved models' of self-dual Yang-Mills and self-dual gravity that uplift naturally to twistor space. In these models, the loop amplitudes vanish via an anomaly cancellation mechanism, analogous to the Green-Schwarz mechanism in superstring theory \cite{Green:1984sg}. Building on these works, ref.~\cite{Monteiro:2022nqt} described the quantum-corrected actions for the usual (`non-improved') theories of self-dual Yang-Mills and self-dual gravity, and described also how the integrability anomaly in these theories may be expressed as the quantum non-conservation of the tower of currents that underlie the classical integrability. The quantum-corrected actions, which are non-local, were shown to admit remarkably simple expressions in light-cone gauge, neatly exhibiting crucial properties of the amplitudes. Surprisingly, it was found that a version of the BCJ double copy for amplitudes \cite{Bern:2008qj,Bern:2019prr} applies directly to the one-loop (loop-integrated) amplitudes; this is in addition to the expected BCJ double copy for loop integrands \cite{Bern:2010ue}, which for the self-dual sectors is well known to be straightforwardly realised in light-cone gauge  \cite{Boels:2013bi,Monteiro:2011pc}.

These recent developments in interpreting the amplitudes in the self-dual theories as being related to an anomaly are in line with prescient old suggestions \cite{Chalmers:1996rq}, particularly regarding a connection to the non-local effective actions that appear in the context of the trace anomaly. More recently, a comment reported as footnote [11] of ref.~\cite{Costello:2022upu} also pointed in this direction.

The trace anomaly, also known as the Weyl or conformal anomaly, represents the failure of theories that are conformally invariant at the classical level to preserve this property at loop level, as revealed by quantum contributions to the trace of the stress-energy tensor. This subject has a long history \cite{Capper:1974ic,Deser:1976yx,Duff:1977ay,Birrell:1982ix,Duff:1993wm}. We are particularly interested in the non-local type of action introduced in \cite{Riegert:1984kt,Fradkin:1983tg} to incorporate this quantum correction; see \cite{Barvinsky:2023exr} for a recent overview of this mechanism and its applications. We will see that this is precisely the type of non-local quantum correction appearing in self-dual gravity and self-dual Yang-Mills theory. In fact, the emphasis on conformal symmetry in the study of the trace anomaly allows us to understand better the conformal properties of the quantum self-dual theories, connecting to the results of \cite{Henn:2019mvc}. Furthermore, the properties of self-dual fields lead to a close relation between the trace anomaly and another important type of anomaly, the chiral (or axial) anomaly, with an even longer history \cite{Adler:1969gk,Bell:1969ts,Bardeen:1969md,Delbourgo:1972xb}; see e.g.~\cite{Jackiw:2008,Fujikawa:2004cx}. Refs.~\cite{Chattopadhyay:2020oxe,Chattopadhyay:2021udc} observed already a close connection between the chiral anomaly and the amplitudes in self-dual Yang-Mills theory, and the chiral anomaly also plays an important role in the twistor constructions of \cite{Costello:2021bah,Bittleston:2022nfr}. We will discuss the connection to the chiral anomaly, which in turn clarifies the structure of the broken classical integrability of the theories.

In recent years there has been great interest in developing new tools for classical and quantum field theory on non-trivial backgrounds, including the application of modern ideas from the study of scattering amplitudes. The self-dual sectors provide a toy model for extremely challenging loop calculations on such backgrounds. In fact, they are not just a toy model, because they correspond to well-defined physical sectors of gravity and Yang-Mills theory. The connection between the one-loop amplitudes in these sectors and the trace anomaly will provide clues for extending the one-loop results to curved spacetimes.

While most of our focus is on the self-dual theories, there is an intriguing relation between the one-loop amplitudes of self-dual gravity and the two-loop ultraviolet divergence of pure general relativity \cite{Bern:2015xsa,Bern:2017puu}, which we will comment on in light of the new understanding of the one-loop amplitudes.

The paper is organised as follows. In section~\ref{sec:trace}, we review the basic aspects of the trace and chiral anomalies. In section~\ref{sec:sdanomaliestr}, we discuss the construction of the quantum-corrected actions for self-dual gravity and self-dual Yang-Mills theory, relating them to recent work on their interpretation as following from an anomaly, and pointing out their connection to the trace anomaly. In section~\ref{sec:sdanomaliesch}, we discuss the origin of the quantum corrections in terms of a U(1) chiral-type anomaly, and the relation of the latter to the breaking of integrability. In addition, we draw attention to the close similarities between the usual chiral anomaly of fermions in two different settings: 4D self-dual Yang-Mills theory and the 2D Schwinger model. In section~\ref{sec:insights}, we return to the trace anomaly context and its lessons for the conformal properties of the quantum self-dual theories, leading to proposals for the quantum-corrected actions on non-trivial backgrounds. In section~\ref{sec:UVgrav}, we will briefly comment on the relationship between the anomaly in self-dual gravity and the two-loop divergence of pure gravity. We conclude with a final discussion in section~\ref{sec:conclusion}.


\section{Review of the trace and chiral anomalies}
\label{sec:trace}

In this section, we will review the trace anomaly and the chiral anomaly. These typically arise in the context of massless fields on non-trivial backgrounds. Massless bosons and fermions play a similar role on self-dual backgrounds, in particular when we consider their integrating out, which is equivalent up to a sign, as we will discuss. Thus, the parallels between trace and chiral anomalies will be further strengthened in the setting of self-dual fields.

\subsection{Trace anomaly and non-local actions}
\label{sec:TraceReview}

In four dimensions, a classically Weyl-invariant theory generically sees this invariance broken by loop effects on a curved spacetime background. The trace of the quantum-corrected (q.c.) stress-energy tensor may be expressed as \cite{Capper:1974ic,Deser:1976yx,Duff:1977ay,Birrell:1982ix,Duff:1993wm}
\[
\label{eq:qctrace}
T_{(\text{q.c.})}{}^\mu{}_\mu=\alpha C^2 + \beta (E - \tfrac23\square R) + \gamma F^2\,,
\]
where we have also allowed for a background Yang-Mills field. We denote $\square=g^{\mu\nu}\nabla_\mu\nabla_\nu$, with $\nabla_{\mu}$ the covariant derivative, and define
\begin{align}
    E &= R_{\mu\nu\lambda\rho}R^{\mu\nu\lambda\rho}-4R_{\mu\nu}R^{\mu\nu}+R^2, \\
    C^2& = C_{\mu\nu\lambda\rho}C^{\mu\nu\lambda\rho}=R_{\mu\nu\lambda\rho}R^{\mu\nu\lambda\rho}-2R_{\mu\nu}R^{\mu\nu}+\frac1{3}R^2, \\
    F^2 &= \Tr F_{\mu\nu}F^{\mu\nu},
\end{align}
where $\sqrt{|g|}E$ is the Euler density, $C_{\mu\nu\lambda\rho}$ is the Weyl tensor, and $F_{\mu\nu}$ is the field strength tensor. The coefficients $\alpha$, $\beta$, $\gamma$ are theory-dependent parameters. The quantum effect represented by this trace anomaly (also called Weyl anomaly or conformal anomaly) can be incorporated into a quantum-corrected action via the addition of a non-local contribution \cite{Riegert:1984kt,Fradkin:1983tg},
\begin{align}
    \label{eq:Stracean}
    \Gamma_\text{tr}[g,A] = \frac{1}{4}\int d^4x \sqrt{|g|}\left( \alpha C^2+\frac{\beta}{2} \Big(E-\frac{2}{3}\square R\Big)+\gamma F^2\right)\frac1{\Delta_4}  \left(E-\frac{2}{3}\square R\right)\,,
\end{align}
such that
\begin{equation}
    \label{TfromS}
    T_{(\text{q.c.})}{}^\mu{}_\mu = \frac{2}{\sqrt{|g|}}\,g_{\mu\nu}\,\frac{\delta}{\delta g_{\mu\nu}}\Gamma_\text{tr}[g,A]\,.
\end{equation}
The non-locality arises from the inverse of the Fradkin-Tseytlin-Paneitz (FTP) operator \cite{Fradkin:1981jc,Fradkin:1982xc,Paneitz_2008}, a fourth-order differential operator given by
\[
\label{eq:Paneitz}
\Delta_4 = \square^2 + 2 R^{\mu\nu}\nabla_\mu\nabla_\nu-\frac{2}{3} R\, \square + \frac1{3} (\nabla^\mu R)\nabla_\mu\,.
\]
This operator is the unique `completion' of $\square^2$ such that $\sqrt{|g|}\Delta_4$ is Weyl invariant, acting on a scalar of zero Weyl weight. The inverse operator acts in the usual manner, via the Green's function $G(x,x')$ defined from $\sqrt{|g|}\,\Delta_4\,G(x,x')=\delta(x,x')$, that is, 
\[
\frac1{\Delta_4}f(x)=\int d^4x' \sqrt{|g|(x')}\,G(x,x')f(x').
\]

The non-local quantum correction  \eqref{eq:Stracean} can be recast as a local action via the introduction of new scalar degrees of freedom \cite{Riegert:1984kt,Fradkin:1983tg,Barvinsky:1994cg,Deser:1996na,Deser:1999zv,Mazur:2001aa,Meissner:2007xv,Coriano:2017mux,Barvinsky:2023exr}. Due to the asymmetry in the factors on either side of the inverse FTP operator, this would generically require two scalar fields. However, because arbitrary Weyl invariant terms can be added to \eqref{eq:Stracean} without modifying the form of the anomaly, it is possible to construct a local formulation of the quantum-corrected action in terms of a single scalar field. To this end, consider adding the following three non-local Weyl-invariant terms to eq.~\eqref{eq:Stracean}:
\begin{equation}
\label{eq:addWeylinv}
    \frac{\alpha^2}{8\beta}\int d^4x\sqrt{|g|} C^2 \frac{1}{\Delta_4}C^2\,, \quad
    \frac{\gamma^2}{8\beta}\int d^4x\sqrt{|g|} F^2 \frac{1}{\Delta_4}F^2\,, \quad
    \frac{\alpha\gamma}{4\beta}\int d^4x\sqrt{|g|} F^2 \frac{1}{\Delta_4}C^2\,.
\end{equation}
The result is a non-local action that is symmetric in the factors on either side of the inverse FTP operator
\begin{align}
    \label{eq:StraceanSym}
    &\Gamma_\text{tr}[g,A] = \nonumber\\
    &\frac{1}{8\beta}\int d^4x \sqrt{|g|}\left( \alpha C^2+\beta \Big(E-\frac{2}{3}\square R\Big)+\gamma F^2\right)\frac1{\Delta_4}  \left( \alpha C^2+\beta \Big(E-\frac{2}{3}\square R\Big)+\gamma F^2\right).
\end{align}
Due to the Weyl invariance of the added terms, eq.~\eqref{TfromS} continues to hold. Given this ambiguity of adding Weyl-invariant terms, the choice \eqref{eq:StraceanSym} is noteworthy because it can be obtained by integrating out a single auxiliary scalar field $\varphi$ from the following local action \cite{Riegert:1984kt,Fradkin:1983tg}:
\begin{equation}
    \label{eq:StraceanLocal}
    \Gamma_\text{tr}[g,A,\varphi] =\int d^4 x \sqrt{|g|}\left(-2\beta'\varphi\Delta_4\varphi+\varphi\left[\alpha' C^2+\beta' \Big(E-\frac{2}{3}\square R\Big) + \gamma' F^2\right]\right).
\end{equation}
The new field $\varphi$ is a dimension-zero scalar, which has previously been referred to as the ``conformalon'', due to the fact that linear shifts in $\varphi$ are related to Weyl transformations of the metric~\cite{Mottola:2016mpl}.\footnote{An alternative scalar action for the Weyl anomaly was used to prove the $a$-theorem in \cite{Komargodski:2011vj}.} As we will discuss in the following section, similar dimension-zero scalars have appeared elsewhere in the literature, relating to anomalies in the self-dual sectors of Yang-Mills theory and gravity. It is important to note that the parameters appear `primed' in \eqref{eq:StraceanLocal} with respect to \eqref{eq:StraceanSym}. If the conformalon is integrated out naively, the coefficients would be the same, but in fact there is a change in the parameters, due to the contribution from the determinant of the operator $\Delta_4$. This determinant affects the measure of the path integral over the metric, which here can be incorporated back into the action \eqref{eq:StraceanSym} as a redefinition of the parameters, possibly up to Weyl-invariant contributions (and we have ignored also a local $R^2$ term in the Lagrangian density) \cite{Riegert:1984kt}.

Notice that non-local actions like  \eqref{eq:StraceanSym} aim to incorporate the quantum effects of the conformal fields into the gravitational action. This is in general a rather complicated problem. However, simple expressions such as \eqref{eq:StraceanSym} are possible because we are looking specifically at contributions that reproduce the trace anomaly, as opposed to those that capture the full quantum effects. One might hope, however, that the effect of integrating out a conformal field is much simpler if we consider a self-dual background. We will see that this is the case in the following sections.

\subsection{Chiral anomaly}
\label{sec:reviewchiral}

The chiral anomaly is the quantum violation of the classical conservation of a chiral current; see \cite{Jackiw:2008} for a basic introduction. The standard example is that of a Dirac fermion coupled to a gauge field,
\[
S=\int d^4x\; i\bar\psi \,\gamma^\mu(\partial_\mu-iA_\mu)\, \psi\,.
\]
The Noether current\, $j^\mu=\bar\psi \gamma^\mu\gamma^5 \psi$\,, associated to the chiral symmetry of the action under $\psi\mapsto e^{i\theta\gamma^5}\psi$, is classically conserved, but this property is anomalous. In the quantum-corrected theory, we have \cite{Adler:1969gk,Bell:1969ts,Bardeen:1969md}
\[
\partial_\mu\, j^\mu_\text{\,(q.c.)} = \frac1{16\pi^2}\, \varepsilon^{\mu\nu\lambda\rho}\, \text{tr}\, F_{\mu\nu}\,F_{\lambda\rho}\,.
\]
Similarly, for a Dirac field minimally coupled to gravity, we have \cite{Delbourgo:1972xb}
\[
\label{eq:fermanomgrav}
\nabla_\mu\, j^\mu_\text{\,(q.c.)} = -\frac1{384\pi^2}\, \varepsilon^{\mu\nu\lambda\rho}\, R_{\mu\nu\alpha\beta}\,R_{\lambda\rho}{}^{\alpha\beta}\,.
\]

Suppose that we are interested in a quantum-corrected action. We want to know what is the effect, say for a gauge theory action, of integrating out the fermion. This is a very challenging problem in general. It turns out to be simple in two spacetime dimensions \cite{Johnson:1963vz,Jackiw:1983nv}, where, as we will review in section~\ref{eq:fermion4D2D}, the answer in the Schwinger model is reminiscent of the gauge field piece in \eqref{eq:StraceanSym}. In four dimensions, we may hope that the problem is relatively simple in the case of self-dual gauge theory, and in fact the aforementioned gauge field piece in \eqref{eq:StraceanSym} will suffice for certain choices of the gauge group.


\section{Anomalies in the self-dual theories and the trace anomaly}
\label{sec:sdanomaliestr}

A recent instance of the appearance of non-local actions associated to anomalies is the quantum description of self-dual gravity (SDG) and self-dual Yang-Mills theory (SDYM). This section is partly a review of these developments, and partly a new perspective on their connection to the trace anomaly. It also sets the stage for the connection to the chiral anomaly, to be discussed in the following section.

SDG and SDYM are well-known to be classically integrable, and have the associated property of vanishing tree-level scattering amplitudes. In fact, the only scattering amplitudes are at one-loop with positive-helicity external states \cite{Bern:1993qk,Mahlon:1993si,Bern:1996ja,Bern:1998xc,Bern:1998sv}. For Yang-Mills theory, we have in spinor-helicity language the following colour-ordered amplitudes \cite{Bern:1993qk}:
\[
\label{eq:An}
A^{(1)}_\text{YM}(1^+2^+\cdots n^+) 
= -\frac{i}{(4\pi)^2\cdot 3}  \sum_{1\leq i_1<i_2<i_3<i_4\leq n} \frac{\langle i_1i_2\rangle [i_2i_3]\langle i_3i_4 \rangle [i_4i_1] }{\langle 12\rangle\langle 23\rangle \cdots \langle n1\rangle}\,,
\]
which correspond to the self-dual sector of the full theory.
For the analogous case in gravity, the $n$-point expression is more involved, but is also known in closed form \cite{Bern:1998xc}. The 4-point case can be written as
\begin{align}
\label{eq:M4}
A^{(1)}_\text{G}(1^+2^+3^+4^+) = -\frac{i}{(4\pi)^2\cdot 120}\, (s_{12}^2 + s_{23}^2 + s_{31}^2)\left(\frac{s_{12}s_{23}}{\langle12\rangle\langle23\rangle\langle34\rangle\langle41\rangle}\right)^2 
\,.
\end{align}
In both SDYM and SDG, the amplitudes are rational functions of the external data, i.e.~they possess poles but not branch cuts, because the tree amplitudes vanish. The simplicity of the amplitudes was argued in \cite{Bardeen:1995gk} to be a remnant of the integrability of the classical theories, with the amplitudes resulting from an anomaly. Recent work, to be discussed in the following, has provided an elegant realisation of this idea. Let us stress that the particular anomaly we are considering here is not a pathological feature of quantum SDYM or SDG. It merely represents the failure of the quantum theories to exhibit the integrability that is a property of the classical theories, as signalled by the non-vanishing one-loop amplitudes.

\begin{figure}[t]
\centering
\includegraphics[width=0.6\textwidth]{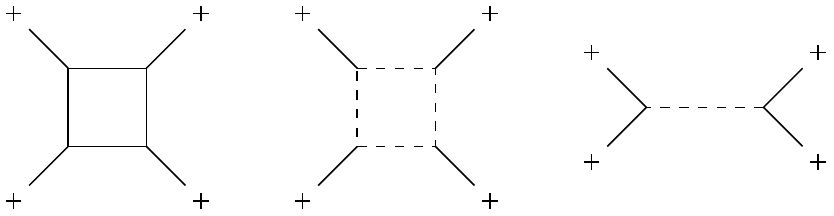}
\caption{The ordinary SDYM loop diagrams with gluons (left) are cancelled in the `improved' theory of ref.~\cite{Costello:2021bah} by tree-level exchanges of the scalar field (right). In the `improved' SDG theory of ref.~\cite{Bittleston:2022nfr}, there is a similar cancellation, involving also an additional type of contribution (centre).}
\label{fig:diagrams}
\end{figure}

Twistor space has long been a convenient setting to study the classical self-dual theories, a fact that is closely related to their integrability; see e.g.~\cite{lionel1996integrability,Dunajski:2010zz,Adamo:2017qyl}. Based on a twistor construction, 
`improved' non-anomalous versions of quantum SDYM and SDG have been introduced recently in \cite{Costello:2021bah,Bittleston:2022nfr}; see also \cite{Costello:2022wso,Costello:2022upu,Bu:2022dis,Bittleston:2022jeq,Bu:2023vjt}. These are improved in the sense that the quantum theories can still be formulated in twistor space, unlike ordinary quantum SDYM and SDG. The resulting  integrability of the `improved' theories is associated to the vanishing of their loop amplitudes. In practice, the loop amplitudes \eqref{eq:An} and \eqref{eq:M4} are cancelled via a Green-Schwarz-type mechanism, analogous to the one in superstring theory \cite{Green:1984sg}: one-loop diagrams of gravitons/gluons are cancelled by diagrams involving an auxiliary axion field. This is represented in Figure~\ref{fig:diagrams}.

We will discuss here the quantum correction to ordinary (i.e.~anomalous) SDYM and SDG, with a particular focus on the similarity of this structure to the trace anomaly. This relies crucially on results presented in \cite{Costello:2021bah,Bittleston:2022nfr}, and we will connect our discussion to the quantum-integrable `improved' versions of SDYM and SDG presented in those works. The logic of our discussion of non-local quantum corrections builds on \cite{Monteiro:2022nqt}.

\subsection{Anomaly in self-dual Yang-Mills theory}

Let us start with SDYM. The usual action is
\[
S_{\text{SDYM}}[B,A]=\int \Tr \, B\wedge F_\text{ASD}\,,
\]
with $B$ a Lagrange-multiplier 2-form that eliminates the anti-self-dual part of the field strength $F$ \cite{Chalmers:1996rq}. The non-vanishing amplitudes in this theory are the all-plus one-loop amplitudes; they correspond to the contributions on the left of Figure~\ref{fig:diagrams}, where all lines are gluons.\footnote{Fadeev-Popov ghosts should be included here if appropriate. They are not required in light-cone gauge, which is particularly convenient for the self-dual theories.} Now, there is a well-known property that these contributions are exactly reproduced by considering a complex minimally-coupled massless scalar in the loop, instead of the gluon. In both cases, there are two real bosonic degrees of freedom running in the loop.\footnote{This is easy to see from the Feynman rules in light-cone gauge. It can also be thought of as a consequence of the so-called supersymmetric Ward identities \cite{Grisaru:1977px,Grisaru:1979re}, which imply that the non-supersymmetric all-plus (and one-minus) amplitudes are proportional to the number of bosonic minus fermionic degrees of freedom running in the loop.} Consider the action
\[
\label{eq:preqcSSDYM}
S_{\text{SDYM}} - \int d^4x\,(D_\mu \vartheta)^\dagger D^\mu \vartheta \,,
\]
like scalar QCD, but with a self-dual gauge field. 
The self-duality condition imposed by the Lagrange-multiplier-type part $S_{\text{SDYM}}$ means that the massless scalar only couples to one helicity of the gluon. The amplitudes for external gluons will receive two contributions: one from gluons in the loop and one from the complex scalar in the loop, with the two contributions being equal, as already mentioned. Therefore, if we integrate out the complex scalar, we obtain an exact effective action for the gluons, with one-loop effective vertices arising from the determinant of the operator acting on the scalar. Integrating out the scalar is a complicated problem on a generic gauge field background, requiring regularisation, but one can hope that it has a simple solution if that background is self-dual. Indeed, in this case, it generates one-loop vertices for UV-finite amplitudes without branch cuts, namely those given in \eqref{eq:An}. Still, a closed-form solution to this problem is at present available only for a particular set of gauge groups, namely SU(2), SU(3), SO(8) or one the exceptional groups; this solution follows from the results of \cite{Costello:2021bah}. The groups mentioned are identified by the fact that the Lie algebra $\mathfrak{g}$ satisfies the following property relating the trace in the adjoint representation to double traces in the fundamental representation:
\[
\Tr_\text{ad}(T^{(a_1}T^{a_2}T^{a_3}T^{a_4)}) = C_\mathfrak{g}\, \Tr(T^{(a_1}T^{a_2})\Tr(T^{a_3}T^{a_4)})\,.
\]
The proportionality constant is 
\[
C_\mathfrak{g} = \frac{10\, ({\bf h}^\vee)^2}{2+ \text{dim}\,\mathfrak{g}}\,,
\]
where ${\bf h}^\vee$ is the dual Coxeter number. For these gauge groups and a self-dual $F$, the results of \cite{Costello:2021bah} imply (we will see later the precise relation) that we obtain the quantum-corrected action
\[
\label{eq:qcSSDYM}
S_{\text{q.c.SDYM}}[B,A] = S_{\text{SDYM}} -\frac{a_\mathfrak{g}^2}{8} \int d^4x\; \Tr(F_{\mu\nu}F^{\mu\nu})
\,\frac1{\square_0^2}\,\Tr(F_{\mu\nu}F^{\mu\nu})\,,
\]
where the subscript in $\square_0$ denotes a flat background and
\[
a_\mathfrak{g} = \sqrt{-\frac{iC_\mathfrak{g}}{12\pi^2}}\,.
\]
For clarity, by quantum-corrected we mean that all loop effects are accounted for explicitly in the action, so that loop amplitudes can be calculated using tree-level-type rules. In particular, the new term introduces one-loop effective vertices.\footnote{This is equivalent to explicitly solving the loop integrals that are usually implicit when incorporating the functional determinants associated to loop effects into an action (see \cite{Lee:2022aiu,Gomez:2022dzk,Kakkad:2022ryl} for recent examples of integrand-level --- rather than integral-level --- approaches).} This simple non-local action produces the expected one-loop amplitudes \eqref{eq:An}, though this is not immediately obvious. Notice that the restriction on the gauge group leads to linear relations between colour traces that would be independent for SU($N$). Hence, it is important to consider the colour-dressed amplitude when comparing with \eqref{eq:An}. 

The quantum correction in \eqref{eq:qcSSDYM} is highly reminiscent of the non-local effective actions designed to incorporate the trace anomaly that were discussed in section~\ref{sec:TraceReview}. To sharpen this connection, consider again the contribution \eqref{eq:StraceanSym}. By restricting now to the special case of a gauge field in flat spacetime, it reduces to 
\[
    \label{eq:tranomGammaYM}
 \Gamma_\text{tr}[g_{\mu\nu}=\eta_{\mu\nu},A] = \frac{\gamma^2}{8\beta}\int d^4x \,\Tr( F_{\mu\nu}F^{\mu\nu})\frac1{\square_0^2} \Tr (F_{\mu\nu}F^{\mu\nu})\,.
\]
We note that on Minkowski spacetime, the FTP operator~\eqref{eq:Paneitz}, whose inverse appears in~\eqref{eq:StraceanSym}, evaluates to $\square_0^2$ in the expression above. This quantum correction is thus identical to the SDYM correction in~\eqref{eq:qcSSDYM}, valid for the restricted choice of gauge groups. Furthermore, the specific form of the trace anomaly quantum correction in~\eqref{eq:StraceanSym} was engineered such that it could be rendered local via the introduction of a dimension-zero scalar field. The general form of this local action was given in~\eqref{eq:StraceanLocal}, and for a gauge field on flat spacetime it reduces to
\begin{equation}
    \label{eq:StraceanLocal1}
    \Gamma_\text{tr}[g_{\mu\nu}=\eta_{\mu\nu},A,\varphi] =\int d^4 x \left(-2\beta\,\varphi\,\square_0^2\,\varphi+\gamma \,\varphi\, \Tr( F_{\mu\nu}F^{\mu\nu}) \right).
\end{equation}
Indeed, integrating out the scalar in this expression yields~\eqref{eq:tranomGammaYM}. Note that in contrast to~\eqref{eq:StraceanSym} and~\eqref{eq:StraceanLocal} there is no need for primed coefficients here. This is due to the fact that the operator $\square_0^2$ is independent of the gauge field, and therefore when integrating out $\varphi$ the effect on the measure of the path integral is trivial.

We can conclude from this discussion that integrating out $\vartheta$ in the action \eqref{eq:preqcSSDYM} has, for the gauge groups considered, precisely the same effect as integrating out $\varphi$ --- which played the role of the conformalon in the trace anomaly --- in the action
\[
\label{eq:confqcSSDYM}
S_{\text{q.c.SDYM}}[B,A,\varphi]=S_{\text{SDYM}} + \int d^4 x \left(\frac1{2}\,\varphi\,\square_0^2\,\varphi+\frac{a_{\mathfrak{g}}}{2} \,\varphi\, \Tr(F_{\mu\nu}F^{\mu\nu}) \right) \,.
\]
Here, as in~\eqref{eq:StraceanLocal1}, integrating out $\varphi$ is a simple algebraic operation, because $\square_0^2$ does not depend on the gauge field. We are now in a position to consider an `improved' model of SDYM, which we may call $\varphi-$SDYM; improved in the sense that it has vanishing amplitudes, and is therefore quantum-integrable. Its (non-quantum-corrected) action is
\[
\label{eq:confintSSDYM}
S_{\varphi-\text{SDYM}}[B,A,\varphi] = S_{\text{SDYM}} + \int d^4 x \left(\frac1{2}\,\varphi\,\square_0^2\,\varphi+\frac{ia_{\mathfrak{g}}}{2} \,\varphi\, \Tr(F_{\mu\nu}F^{\mu\nu}) \right) \,.
\]
Due to the change of coefficient in the last term, with respect to \eqref{eq:confqcSSDYM}, integrating out $\varphi$ now leads to \eqref{eq:qcSSDYM} with a different sign for the last term. That is, the effect of the scalar is now to cancel the one-loop gluon diagrams arising from $S_{\text{SDYM}}$. In terms of the Figure~\ref{fig:diagrams}, the gluon diagrams on the left are cancelled by the diagrams on the right, with tree exchanges of $\varphi$. The diagrams in the centre of Figure~\ref{fig:diagrams} are absent for the action \eqref{eq:confintSSDYM}, which is why integrating out $\varphi$ is straightforward. There is one last step before recognising the equivalence of this `improved' theory of SDYM to the one obtained in \cite{Costello:2021bah} from a twistor-space construction. That is to notice that the Lagrange-multiplier-type action $S_{\text{SDYM}}$ is imposing self-duality of the gauge field, such that
\[
\label{eq:SDFF} \Tr(F_{\mu\nu}F^{\mu\nu})=-\frac{i}{2}\Tr(\varepsilon^{\mu\nu\rho\lambda}F_{\mu\nu}F_{\rho\lambda})\,. 
\]
Hence,~\eqref{eq:confintSSDYM} can be written equivalently as 
\[
\label{eq:axionintSSDYM}
S_{\varphi-\text{SDYM}}[B,A,\varphi] = S_{\text{SDYM}} + \int  \left(d^4 x\;\frac1{2}\,\varphi\,\square_0^2\,\varphi+a_{\mathfrak{g}} \,\varphi\Tr (F\wedge F) \right) \,.
\]
This is the theory introduced in \cite{Costello:2021bah}. The scalar $\varphi$ has an axionic coupling to the gauge field here, whereas it had a `dilatonic' coupling in \eqref{eq:confintSSDYM}. This ambiguity arises naturally from the self-duality condition.

It would be interesting to work out explicitly the extension of this story to SU($N$) SDYM. The first step in this direction, reproducing the parity-even part of the quantum correction to standard SDYM, i.e.~generalising \eqref{eq:qcSSDYM}, was taken in \cite{Monteiro:2022nqt}. This functional corresponds to
\[
\label{eq:Epart}
\int d^4x\; \left(\frac1{D^2}(F_{\mu\nu}F^{\mu\nu})\right)^{ab} \left(\frac1{D^2}(F_{\mu\nu}F^{\mu\nu})\right)^{ab}\,,
\]
where $a,b$ are Lie algebra indices; see \cite{Monteiro:2022nqt} for the definition of the $1/D^2$ operation.
Still missing in closed form is a parity-odd part carrying a Levi-Civita symbol $\varepsilon^{\mu\nu\alpha\beta}$, which contributes to scattering amplitudes at multiplicity $n>4$.

Let us return to the question of the similarity between the quantum correction in \eqref{eq:qcSSDYM} and the type of non-local expression found in the context of the Weyl anomaly. This does not mean that the anomaly is of Weyl type. The appearance of the functional in the Weyl context came via the introduction in \eqref{eq:addWeylinv} --- in particular, the term in the centre --- of Weyl-invariant terms to complete the perfect square in \eqref{eq:StraceanSym}. The anomaly in SDYM is actually related to the breaking of a chiral symmetry, as we will discuss in section~\ref{eq:fermion4D2D}. Interestingly, refs.~\cite{Chattopadhyay:2020oxe,Chattopadhyay:2021udc} presented a construction analogous to the chiral anomaly that reproduces the loop amplitudes for SU($N$). This construction was based in `region-momenta' space (adapted to planar amplitudes) instead of position space. One drawback is that the application of a similar construction to gravity (which has no planar sector) is unclear. 

Notice that the quantum correction in \eqref{eq:qcSSDYM} can also be seen as the result of integrating out a massless fermion --- up to a sign. Indeed, a well-known procedure to obtain an `improved' SDYM theory, alternative to the bosonic model with the axion introduced in \cite{Costello:2021bah}, is to introduce supersymmetry. The sign of the quantum correction in \eqref{eq:qcSSDYM} being flipped can be achieved by integrating out a fermion with two real degrees of freedom (the gaugino), instead of integrating out the scalar in \eqref{eq:confintSSDYM} or \eqref{eq:axionintSSDYM}. All-plus amplitudes are absent in super-Yang-Mills theories, and therefore also in their self-dual sectors; the same property applies to all-plus amplitudes of gravitons in supergravity theories (and to one-minus amplitudes in both cases).

While we will later discuss the interpretation of the quantum correction as resulting from a chiral anomaly, thinking about the trace anomaly is helpful too, because it is better studied from the point of view of non-local functionals, particularly regarding the appearance of a fourth-derivative Weyl-covariant operator that will be useful in later sections.

\subsection{Anomaly in self-dual gravity}

The story for SDG is broadly similar to that of SDYM. The usual action is
\[
S_{\text{SDG}}[\Gamma,e]=\int \Gamma_{\alpha\beta}\wedge d(e^{\alpha\dot\alpha}\wedge e^\beta{}_{\dot\alpha})\,,
\]
where we have $ds^2=\epsilon_{\alpha\beta}\epsilon_{\dot\alpha\dot\beta}e^{\alpha\dot\alpha}e^{\beta\dot\beta}$, and the three 1-forms $\Gamma_{\alpha\beta}=\Gamma_{\beta\alpha}$ play the role of Lagrange multipliers \cite{Plebanski:1977zz,Ashtekar:1987qx,Smolin:1992wj,Capovilla:1991qb}.\footnote{Plebanski's two heavenly equations for SDG \cite{Plebanski:1975wn} follow from this action: (i) the Lagrange multipliers enforce that the three self-dual 2-forms $e^{\alpha\dot\alpha}\wedge e^\beta{}_{\dot\alpha}$ are closed, and (ii) each of the heavenly equations corresponds to a choice of Darboux coordinates allowed by that closure; see e.g.~\cite{Adamo:2021bej} for details. See also \cite{Krasnov:2021cva} for other choices of SDG action.} The non-vanishing amplitudes in this theory are the all-plus one-loop amplitudes; they correspond to the type of diagram on the left of Figure~\ref{fig:diagrams}, where all lines are gravitons. Just as in the SDYM case, these amplitudes are exactly reproduced by having a complex minimally-coupled massless scalar running in the loop, instead of the graviton; both have two bosonic real degrees of freedom. Consider the action
\[
\label{eq:preqcSSDG}
S_{\text{SDG}} + \int d^4x\,\sqrt{|g|}\; \bar\sigma\,\square\,\sigma \,.
\]
The self-duality condition imposed by $S_{\text{SDG}}$ means that the massless scalar only couples to one helicity of the graviton.
The amplitudes for external gravitons receive two equal contributions, corresponding to the graviton or the complex scalar running in the loop.\footnote{The dilaton and axion fields of `$\mathcal N=0$ supergravity' would be another way to provide an equal contribution to that of gravitons in the loop.} Therefore, if we integrate out the complex scalar, we obtain an exact quantum-corrected action for the gravitons, with one-loop effective vertices arising from the determinant of the scalar kinetic operator. A consequence of the results of \cite{Bittleston:2022nfr} (we will see later the precise relation) is that the quantum-corrected action for SDG can be written as
\[
\label{eq:qcSSDG}
S_{\text{q.c.SDG}}[\Gamma,e] = S_{\text{SDG}} -\frac{a_\text{G}^2}{16} \int d^4x\,\sqrt{|g|}\; C^2 \,\frac1{\square^2}\,C^2 \,,
\]
where
\[
a_\text{G} = \sqrt{-\frac{i}{960\pi^2}}\,.
\]

The quantum correction is again reminiscent of that encountered in the context of the trace anomaly. Indeed, considering the trace anomaly functional~\eqref{eq:StraceanSym} for a vanishing gauge field and a Ricci-flat metric (because self-dual metrics are Ricci-flat), one obtains
\[
 \label{eq:StraceanLocalG}
 \Gamma_\text{tr}[g|_{R_{\mu\nu}=0},A=0] = \frac{(\alpha+\beta)^2}{8\beta}\int d^4x \sqrt{|g|}\, C^2\frac1{\square^2}C^2\,.
\]
This reproduces precisely the quantum correction for SDG given in~\eqref{eq:qcSSDG}. Furthermore, as discussed in section~\ref{sec:TraceReview}, the trace anomaly functional can be obtained by integrating out the scalar $\varphi$ from a local action~\eqref{eq:StraceanLocal}, which in this case corresponds to
\begin{equation}
    \label{eq:StraceanLocal2}
    \Gamma_\text{tr}[g|_{R_{\mu\nu}=0},\varphi] =\int d^4 x \sqrt{|g|} \left(-2\beta'\,\varphi\,\square^2\,\varphi+(\alpha'+\beta')\, \varphi\, C^2 \right)\,.
\end{equation}
Here, when integrating out $\varphi$, the determinant of the operator $\square^2$ affects the measure of the path integral, an effect which is taken into account by a redefinition of the parameters, as indicated by primes. 

In light of this close relation between the quantum correction for SDG and that incorporating the trace anomaly, let us discuss further the integrating out of the complex scalar from \eqref{eq:preqcSSDG} to \eqref{eq:qcSSDG}. We can consider in \eqref{eq:preqcSSDG} the addition of the following 2-point vertex for the scalar,
\[
\label{eq:preqcSSDG2}
S_{\text{SDG}} + \int d^4x\,\sqrt{|g|}\Big(\bar\sigma\,\square\,\sigma -\frac1{2} \bar\sigma^2\Big)\,.
\]
Importantly, on the support of the self-duality condition imposed by the Lagrange-multiplier-type action $S_{\text{SDG}}$, this does not introduce a new interaction with the metric (this can be seen in light-cone gauge, where $g$ is a constant) \cite{Bittleston:2022nfr}. Integrating out $\bar\sigma$, we obtain
\[
\label{eq:preqcSSDG3}
S_{\text{SDG}} + \int d^4x\,\sqrt{|g|}\;\frac1{2}\,\sigma\,\square^2\,\sigma\,.
\]
Therefore, integrating out the complex scalar, from \eqref{eq:preqcSSDG} to \eqref{eq:qcSSDG}, is equivalent to integrating out the `conformalon'-like scalar $\sigma$, from \eqref{eq:preqcSSDG3} to \eqref{eq:qcSSDG}.\footnote{Notice that we have, by the procedure \eqref{eq:preqcSSDG2}, assigned a mass dimension of 2 to $\bar\sigma$ and 0 to $\sigma$.} This equivalence is also consistent with the fact that, on a self-dual background, these fields give precisely the same contribution to the trace anomaly. The contributions to the trace \eqref{eq:qctrace} are $(\alpha,\beta)=\frac1{(4\pi)^2}(\frac1{60},-\frac1{180})$ for the complex minimally-coupled scalar, which has two real degrees of freedom, and $(\alpha,\beta)=\frac1{(4\pi)^2}(-\frac1{15},\frac{7}{90})$ for the scalar with fourth-derivative kinetic term (see e.g.~\cite{Antoniadis:1992xu} for the coefficients used here). Self-dual backgrounds are Ricci-flat, and therefore we have $E-\frac{2}{3}\square R=C^2$. In both cases, we obtain $\alpha+\beta=\frac1{(4\pi)^2 \; 90}\,,$ so their contributions to the trace anomaly are equivalent. From the discussion of the trace anomaly above, particularly the integrating out of the conformalon, it is thus natural to expect the functional appearing in the quantum correction in \eqref{eq:qcSSDG}. The coefficient of that quantum correction cannot, however, be determined from the trace anomaly, because the effective action for the latter is only defined up to Weyl-invariant terms, which can take the form of the sought functional --- see the first term in \eqref{eq:addWeylinv}. Instead, the coefficient in \eqref{eq:qcSSDG} can be fixed by comparing to a one-loop amplitude of SDG, say at 4 points. This is, however, an indirect way of computing the quantum correction to the SDG action, or at least of finding a natural candidate and checking the result.

We may proceed, as for SDYM, to obtain an `improved' quantum-integrable upgrade of SDG, with vanishing amplitudes; we call it $\varphi$-SDG. The idea is to flip the sign of the quantum correction in \eqref{eq:qcSSDG}, and interpret that as a (non-quantum-corrected) action, where the last term results from integrating out a scalar field. Given that we know the effect of integrating out $\sigma$ in \eqref{eq:preqcSSDG3}, we can conclude that the following action works,
\[
\label{eq:confintSSDG}
S_{\varphi-\text{SDG}}[\Gamma,e,\varphi] = S_{\text{SDG}} + \int d^4 x \sqrt{|g|}\left(\;\frac1{2}\,\varphi\,\square^2\,\varphi - \frac{ia_\text{G}}{2}\, \varphi\, C^2 \right)\,.
\]
We called the scalar $\varphi$ to emphasise the similarity to the `improved' SDYM action \eqref{eq:confintSSDG} and to the conformalon of the trace anomaly. The amplitudes of gravitons in this action vanish as a cancellation of the contributions represented in Figure~\ref{fig:diagrams}: on the left, the loops diagrams of gravitons from $S_{\text{SDG}}$; in the centre, the loop diagrams with $\varphi$ running in the loop, resulting from the scalar-graviton interactions included in its kinetic term; and on the right, the tree exchanges of $\varphi$, resulting from the last term in \eqref{eq:confintSSDG}. We have now the theory introduced in \cite{Bittleston:2022nfr} from a twistor-space construction. This becomes clear when we note that, on the self-duality constraint imposed by $S_{\text{SDG}}$, we have
\[
C^2 = \frac{i}{2}\varepsilon^{\rho\lambda\sigma\omega}R^\mu{}_{\nu\rho\lambda} \,R^\nu{}_{\mu\sigma\omega}\,.
\]
Hence, we can write the theory with the axionic-like coupling as in \cite{Bittleston:2022nfr},
\[
\label{eq:axionintSSDG}
S_{\varphi-\text{SDG}}[\Gamma,e,\varphi] = S_{\text{SDG}} + \int \left( d^4 x \sqrt{|g|}\;\frac1{2}\,\varphi\,\square^2\,\varphi + a_\text{G}\, \varphi\, \mathsf{R}^\mu{}_\nu\wedge \mathsf{R}^\nu{}_\mu \right)\,.
\]

Just as we mentioned for SDYM, an alternative to the bosonic model of \cite{Bittleston:2022nfr} is to cancel the one-loop amplitudes of SDG using a massless fermion with two real degrees of freedom --- as in supergravity. Integrating out this fermion then leads precisely to the second term in \eqref{eq:qcSSDG}, with the opposite sign. Furthermore, as was the case for SDYM, there is an interpretation of the anomaly as a chiral anomaly, as we will soon discuss.

Just before the submission of our paper, ref.~\cite{Karateev:2023mrb} appeared, discussing the application of the trace anomaly to graviton-dilaton amplitudes. There may be a connection to our discussion. Finally, we note that a relation between conformal self-dual gravity and the Weyl anomaly was pointed out in \cite{Schmidhuber:1992dg}. It would be interesting to understand how this connects to our discussion for ordinary self-dual gravity.

\subsection{Anomalies in self-dual Einstein-Yang-Mills theory}

For completeness, we can briefly discuss the coupling of SDYM to SDG. The action of self-dual Einstein-Yang-Mills theory (SDEYM) is simply
\[
S_{\text{SDEYM}}[\Gamma,e,B,A]=\int \left(\, \Gamma_{\alpha\beta}\wedge d(e^{\alpha\dot\alpha}\wedge e^\beta{}_{\dot\alpha})
+ B\wedge F_\text{ASD}\,\right),
\]
where the second term depends on the metric via the Hodge dual in the self-duality condition.
Ref.~\cite{Bittleston:2022nfr} introduced the `improved' theory: 
\[
\label{eq:SrhoSDEYM}
S_{\varphi-\text{SDEYM}} = S_{\text{SDEYM}} + \int d^4 x \sqrt{|g|}\,\,\frac{1}{2}(\square\,\varphi)^2 + \int \varphi\, \big(a_\mathfrak{g}  \Tr{F\wedge F} + a_{\text{G},\mathfrak{g}} \,\mathsf{R}^\mu{}_\nu\wedge \mathsf{R}^\nu{}_\mu\big)\,,
\]
where $a_{\text{G},\mathfrak{g}}=a_\text{G}\sqrt{1+\frac1{2}\,\text{dim}\,\mathfrak{g}}$. Here, $\text{dim}\,\mathfrak{g}$ is the dimension of the Lie algebra, with the gauge group still obeying the restriction mentioned above for SDYM. This modification of the coupling $a_\text{G}$ of $\varphi-\text{SDG}$ occurs due to the need to cancel diagrams with external gravitons that have gluons running in the loop. Along the lines of the SDYM and SDG cases, the action above implies that we can write down a quantum-corrected action for the ordinary SDEYM theory as
\begin{align}
\label{eq:qcSSDEYM}
S_{\text{q.c.SDEYM}} = &\;  S_{\text{SDEYM}} 
 \nonumber\\
& - \int d^4x\,\sqrt{|g|}\,\Big( \frac{a_\mathfrak{g}^2}{8} \,F^2
\,\frac1{\square^2}\,F^2
 + \frac{\tilde a_{\text{G},\mathfrak{g}}^2}{8}\; C^2 \,\frac1{\square^2}C^2 
 \;-\; \frac{a_{\text{G},\mathfrak{g}}\, a_\mathfrak{g}}{4} \; C^2 \frac1{\square^2}\,F^2\,\Big) \,,
\end{align}
where $\,F^2=\Tr(F_{\mu\nu}F^{\mu\nu})$\,.
The only subtlety here is the substitution of the coefficient \,$a_{\text{G},\mathfrak{g}}=a_\text{G}\sqrt{1+\frac1{2}\,\text{dim}\,\mathfrak{g}}$\, by the coefficient \,$\tilde a_{\text{G},\mathfrak{g}}=a_\text{G}\sqrt{\frac1{2}+\frac1{2}\,\text{dim}\,\mathfrak{g}}$\, in the second line. This arises from the functional determinant when integrating out $\varphi$, as explained in the SDG case. This quantum-corrected action reproduces the one-loop scattering amplitudes of self-dual Einstein-Yang-Mills theory. For a brief discussion of how the amplitudes in the self-dual sector relate to those in the full Einstein-Yang-Mills theory \cite{Faller:2018vdz,Nandan:2018ody}, see \cite{Bittleston:2022nfr}; unlike the cases of SDYM and SDG, the self-dual sector does not produce the complete all-plus one-loop amplitudes, but only the piece of lowest order in the gravitational coupling.

\subsection{Quantum-corrected actions in light-cone gauge}
\label{ref:LCactions}

In preparation for the next section, we review here the light-cone formulation of SDYM and SDG. The fact that the quantum-corrected actions, namely \eqref{eq:qcSSDYM} and \eqref{eq:qcSSDG}, lead to loop amplitudes that are non-vanishing only for one-loop and for all-plus external states may not be immediately obvious from those expressions. This is, however, clearly seen in light-cone gauge \cite{Monteiro:2022nqt}. We just quote the results here for illustration, and refer the reader to that work for further details. Consider double-null coordinates such that 
\[
ds^2_0=2(-du dv+dw d\wb)\,,\qquad \square_0=2(-\partial_u\partial_v+\partial_w\partial_\wb)\,,
\]
and define
\[
\buildrel{\leftrightarrow} \over {P} \; = \;
\buildrel{\leftarrow} \over{\partial}_u
\buildrel{\rightarrow} \over{\partial}_w-
\buildrel{\leftarrow} \over{\partial}_w
\buildrel{\rightarrow} \over{\partial}_u\,,
\]
where the null plane $(u,w)$ is picked out by the gauge choice. Then,
the actions for a particular definition of off-shell positive-helicity fields ($\Psi$ and $\phi$) and negative-helicity fields ($\bar\Psi$ and $\bar\phi$) can be written as
\[
\label{eq:SDYMLC}
S_{\text{q.c.SDYM}} = \int d^4 x \;\;\left( \text{tr}\big( \bar\Psi \big(\square_0 \Psi + i\, ( \Psi \Pd \Psi))\big) - \frac{a_\mathfrak{g}^2}{32} \Big(\, \frac1{\square_0} \,\text{tr}( \Psi \Pd\!{}^2\, \Psi ) \Big)^2\right)
\]
and
\[
\label{eq:SDGLC}
S_\text{q.c.SDG} = \int d^4 x \; \left(\,\bar\phi \Big(\square_0 \phi - \frac1{2}(\phi \Pd\!{}^2\, \phi)\Big) -\frac{a_\text{G}^2}{16} \left(\, \frac1{ \square_0 -(\phi  \Pd\!{}^2\;\cdot) }\, (\phi \Pd\!{}^4\, \phi )\,\right)^2\,\right)\,.
\]
The Feynman rules for these actions make it clear that the only loop-level diagrams that can be constructed are those at one loop with external positive-helicity fields.

Notice that the presence of $\frac1{\square^2}$ in the actions above does not lead to unphysical properties. It is actually essential to recover the one-loop amplitudes \eqref{eq:An} and \eqref{eq:M4}. These have at most simple poles $\frac1{p^2}$ in the kinematic data, instead of the poles $\frac1{p^4}$ naively associated to $\frac1{\square^2}$. The reduction of the order of the pole is due to kinematic identities. In the long history of the trace anomaly, there have been doubts about the suitability of $\frac1{\square^2}$-type operations that appear in the non-local effective action, as reviewed in \cite{Barvinsky:2023exr}. In the present instance, at least, such operations are perfectly legal.

This formulation also makes it easier to understand the soft and collinear behaviour of the amplitudes from an action/Feynman-rules perspective. Such behaviour has been recently explored in the context of celestial holography, in terms of the celestial soft operator-product-expansion algebras; see e.g.~\cite{Guevara:2021abz,Strominger:2021lvk,Jiang:2021ovh,Jiang:2021csc,Adamo:2021lrv,Ball:2021tmb,Mago:2021wje,Ren:2022sws,Monteiro:2022lwm,Costello:2022jpg,Bu:2022iak,Bhardwaj:2022anh,Guevara:2022qnm,Ball:2022bgg,Adamo:2022lah,Monteiro:2022xwq,Garner:2023izn,Ren:2023trv,Bittleston:2023bzp,Costello:2023hmi,He:2023lvk,Brown:2023zxm,Taylor:2023ajd,Himwich:2023njb,Chattopadhyay:2024kdq}. Historically, it was precisely this special behaviour that led to the $n$-point formulas for the all-plus one-loop amplitudes, obtained in \cite{Bern:1993qk,Bern:1998xc} via a kind of `soft and collinear bootstrap'.

It is worth noting that all interactions in \eqref{eq:SDYMLC} and \eqref{eq:SDGLC}, including the one-loop vertices, exhibit the appearance of the `kinematic algebra' of the self-dual sector \cite{Monteiro:2011pc}, which underlies the BCJ colour-kinematics duality and double copy \cite{Bern:2008qj} in the case of the self-dual theories.\footnote{It would be interesting to see whether the explicit quantum corrections fit nicely into the formalisms of \cite{Borsten:2022vtg,Borsten:2023paw,Bonezzi:2023pox}, which also aim to expose the kinematic algebra in the self-dual theories.} The antisymmetrised derivatives with respect to $u$ and $w$ lead in Fourier space to the appearance of the structure constant
\[
X(k_1,k_2)=k_{1w}k_{2u}-k_{1u}k_{2w}
\]
of the Lie algebra of area-preserving diffeomorphisms:
\[
[L_{k_1},L_{k_2}]= X(k_1,k_2)\,L_{k_1+k_2}\,,\qquad
L_{k}:=(e^{i\,k\cdot x}\Pd\! \;\cdot\;)\,.
\]
This Lie algebra is equivalent, after a change of basis corresponding to a soft expansion \cite{Monteiro:2022nqt}, to the wedge subalgebra of the Lie algebra of $w_{1+\infty}$, which was identified in the context of celestial holography in \cite{Strominger:2021lvk}. In the celestial OPEs, one factor of the $X$ structure constant enforces the chiral nature of the OPE of positive-helicity gluons or gravitons, whereas only for gravity there is an additional factor of $X$ in the 3-point vertex that leads to the celestial $w_{1+\infty}$ algebra \cite{Monteiro:2022nqt}. So the $(++-)$ `tree' vertex of self-dual gravity, which is picked up in full non-chiral gravity by taking two positive-helicity gravitons, manifestly exhibits the $w_{1+\infty}$ algebra in light-cone gauge. Natural generalisations of this algebra, via deformations or extensions to theories with higher (spacetime) spins, follow in a simple manner from light-cone gauge formulations \cite{Monteiro:2022xwq}. Very recently, ref.~\cite{Himwich:2023njb} presented the first study of $w_{1+\infty}$ in the context of gravity with minimally-coupled massive particles. The light-cone gauge is insightful also in this context, because of the universal appearance of the algebra $X$ in soft factors associated to a positive-helicity graviton. In particular, for a positive-helicity graviton polarisation tensor, we have the appearance of $\,\varepsilon^{+}_{\mu\nu}(k)p^\mu p^\nu=-\frac1{k_u^2}\,X(k,p)^2$\, in the soft factors, where $k$ is the soft graviton momentum, and $p$ is the momentum of an external hard particle, be it massless or massive.


\section{Anomalies in the self-dual theories as chiral anomalies}
\label{sec:sdanomaliesch}

In this section, we describe how the one-loop amplitudes in the self-dual theories can be interpreted as resulting from a chiral anomaly. In fact, we find two distinct interpretations, to be described in the following two subsections. The first is that the broken chiral symmetry pertains to the gluon or graviton fields in the self-dual theories, which is related to the `electric-magnetic' duality of the linearised fields. This chiral anomaly is associated to an infinite tower of anomalous currents signalling the breaking of integrability. The second is that the broken chiral symmetry pertains to a fermion coupled to the Yang-Mills field or the metric, which is the more conventional example of chiral anomaly. Here, the quantum correction arises from integrating out the fermion, and encodes the usual fermionic chiral anomaly. We find an interesting analogy to the 2D Schwinger model.

\subsection{From a chiral anomaly to an integrability anomaly}
\label{ref:chiral}

Here, we describe an understanding of the breaking of integrability as a result of a chiral-type anomaly. This provides a simple answer to the question first posed by Bardeen \cite{Bardeen:1995gk} (see also \cite{Cangemi:1996rx}) of how to construct an infinite tower of currents beyond the classical theory, whose anomaly leads to the one-loop amplitudes.

Let us focus on SDG for our argument. As reviewed previously, the light-cone action is
\[
\label{eq:SDGLCtree}
S_\text{SDG} = \int d^4 x \; \bar\phi \Big(\square_0 \phi - \frac1{2}(\phi \Pd\!{}^2\, \phi)\Big)\,.
\]
Consider the U(1) transformation
\[
(\phi,\bar\phi) &\mapsto (e^{i\theta}\phi,e^{-i\theta}\bar\phi)\,.
\]
This is a chiral-type transformation, acting on the two degrees of freedom according to their helicity. The linearised theory is manifestly invariant under this symmetry, which can be checked to correspond to the well-known `electric-magnetic' duality of linearised gravity, first mentioned in \cite{Penrose:1960eq}. In the non-linear full (i.e.~beyond self-dual) gravity theory, this symmetry is broken by the interactions, as discussed recently in \cite{Monteiro:2023dev}, which builds on the amplitudes literature \cite{Rosly:2002jt,Carrasco:2013ypa,Bern:2013uka,Bern:2017rjw,Novotny:2018iph,Elvang:2019twd,Elvang:2020kuj,Pavao:2022kog,Carrasco:2022jxn,Carrasco:2023qgz}. In the present work, however, we are dealing with the self-dual truncation of gravity. On the one hand, the action above is not invariant under this symmetry due to the 3-point vertex. On the other hand, the tree amplitudes vanish, so at tree-level the amplitudes trivially obey helicity conservation, which states that the only allowed amplitudes have equal number of external plus and minus helicity particles (seen as all incoming, say). If we start from the Feynman rules of \eqref{eq:SDGLCtree}, the vanishing of the tree amplitudes follows from non-trivial identities on the support of on-shell kinematics. However, there is a straightforward alternative way to arrive at this result, which is to consider the non-local field redefinition
\[
(\tilde\phi,\bar{\tilde\phi}) &=\Big(\phi-\frac{1}{2\,\square_0}(\phi \Pd\!{}^2\, \phi),\bar\phi\Big),
\label{eq:phired}
\]
which leads to a free action,
\[
\label{eq:SDGLCtreetilde}
S_\text{SDG} = \int d^4 x \; \bar{\tilde\phi}\, \square_0\, \tilde\phi\,.
\]
Now, the chiral symmetry
\[
(\tilde\phi,\bar{\tilde\phi}) &\mapsto (e^{i\theta}\tilde\phi,e^{-i\theta}\bar{\tilde\phi})
\]
is manifest. The corresponding Noether current is
\[
j_\mu = \bar{\tilde\phi}\; {\buildrel{\leftrightarrow} \over {\partial_\mu}}\; \tilde\phi = \bar{\tilde\phi}\,(\partial_\mu \tilde\phi) - (\partial_\mu \bar{\tilde\phi})\, \tilde\phi\,.
\]
In fact, as expected from the fact that the action is free, we have an infinite number of conserved spin-$s$ currents:
\[
j_{\mu_1\mu_2\cdots\mu_s} = \bar{\tilde\phi}\; {\buildrel{\leftrightarrow} \over {\partial_{\mu_1}}}{\buildrel{\leftrightarrow} \over {\partial_{\mu_2}}}\cdots {\buildrel{\leftrightarrow} \over {\partial_{\mu_s}}} \; \tilde\phi \,,
\label{eq:currents}
\]
such that
\[
\partial^{\mu_1} j_{\mu_1\mu_2\cdots\mu_s} = 
\bar{\tilde\phi}\; {\buildrel{\leftrightarrow} \over {\partial_{\mu_2}}}\cdots {\buildrel{\leftrightarrow} \over {\partial_{\mu_s}}} \, (\square_0\,\tilde\phi) -
(\square_0\,\bar{\tilde\phi})\, {\buildrel{\leftrightarrow} \over {\partial_{\mu_2}}}\cdots {\buildrel{\leftrightarrow} \over {\partial_{\mu_s}}} \; \tilde\phi
\,,
\]
which vanishes according to the equations of motion. The conservation of an infinite tower of currents is a hallmark of integrability, as is the vanishing of the amplitudes.\footnote{The reader familiarised with the amplitudes literature may wonder about the fate of the 3-point $(++-)$ tree amplitudes. These have no support on on-shell real kinematics in Lorentzian signature. They would arise beyond this real Lorentzian setting because $\frac1{\square_0}$ diverges in \eqref{eq:phired} for 3-point on-shell kinematics.}

It was first suggested in \cite{Brandhuber:2006bf} that the field redefinition \eqref{eq:phired} (actually, its SDYM counterpart, but the story is similar) leads to a Jacobian in the path integral measure that generates the one-loop amplitudes. This Jacobian has never been computed from first principles starting from the measure, but we can now identify its explicit form. When the (logarithm of the) Jacobian is incorporated into a quantum-corrected action, we have
\[
\label{eq:SDGLCV}
S_\text{q.c.SDG} = \int d^4 x \; \left(\,\bar{\tilde\phi}\, \square_0\, \tilde\phi + V_{\text{1-loop}}(\tilde\phi)\right)\,.
\]
From \eqref{eq:SDGLC}, we see that we can identify
\[
\label{eq:1loopinttilde}
V_{\text{1-loop}}(\tilde\phi) =  -\frac{a_\text{G}^2}{16} \left(\, \frac1{ \square_0 -(\phi  \Pd\!{}^2\;\cdot) }\, (\phi \Pd\!{}^4\, \phi )\,\right)^2 \Bigg|_{\phi=\phi(\tilde\phi)}\,,
\]
where the right-hand side is evaluated on $\phi=\phi(\tilde\phi)$, the inverse map of the redefinition \eqref{eq:phired}. The Feynman rules for \eqref{eq:SDGLCV} are such that the propagator cannot connect any of the vertices, since these only involve $\tilde\phi$, so the one-loop amplitudes are evaluated at $n$ points solely from the $n$-point vertex\footnote{The effect of substituting $\phi=\phi(\tilde\phi)$ in \eqref{eq:1loopinttilde} is to dress the `old' one-loop vertices with the trees built from the `old' 3-point vertex, so that the `new' vertices \eqref{eq:nptvertex} include all possible contributions at $n$ points.}
\[
\label{eq:nptvertex}
\frac{\delta^n V_{\text{1-loop}}(\tilde\phi)}{\delta \tilde\phi(k_1) \delta \tilde\phi(k_2)\cdots \delta \tilde\phi(k_n)}\Bigg|_{\tilde\phi=0}\,.
\]
Now, we note that the currents \eqref{eq:currents} are anomalous, because the quantum-corrected equations of motion lead to
\begin{align}
\partial^{\mu_1} j_{\mu_1\mu_2\cdots\mu_s} & = 
\bar{\tilde\phi}\; {\buildrel{\leftrightarrow} \over {\partial_{\mu_2}}}\cdots {\buildrel{\leftrightarrow} \over {\partial_{\mu_s}}} \, (\square_0\,\tilde\phi) -
(\square_0\,\bar{\tilde\phi})\, {\buildrel{\leftrightarrow} \over {\partial_{\mu_2}}}\cdots {\buildrel{\leftrightarrow} \over {\partial_{\mu_s}}} \; \tilde\phi \nonumber \\
& = \frac{\delta V_{\text{1-loop}}(\tilde\phi)}{\delta \tilde\phi}
\; {\buildrel{\leftrightarrow} \over {\partial_{\mu_2}}}\cdots {\buildrel{\leftrightarrow} \over {\partial_{\mu_s}}} \; \tilde\phi
\,.
\label{eq:infanom}
\end{align}
Therefore, the integrability symmetry associated to the infinite tower of classically conserved currents is broken by the quantum correction. This is signalled by the non-vanishing of the one-loop amplitudes obtained from \eqref{eq:nptvertex}.

The SDYM case is entirely analogous, based on the field redefinition introduced in \cite{Brandhuber:2006bf}:
\[
(\tilde\Psi,\bar{\tilde\Psi}) &=\Big(\Psi+\frac{i}{\square_0}( \Psi \Pd \Psi),\bar\Psi\Big)\,.
\]
The infinite tower of classically conserved currents is now
\[
\text{tr} \Big( \bar{\tilde\Psi}\; {\buildrel{\leftrightarrow} \over {\partial_{\mu_1}}}{\buildrel{\leftrightarrow} \over {\partial_{\mu_2}}}\cdots {\buildrel{\leftrightarrow} \over {\partial_{\mu_s}}} \; \tilde\Psi \Big)\,.
\]

In either SDG or SDYM, the field redefinition to a free action brings a very simple perspective to the breaking of integrability. A different spacetime-based approach was taken in ref.~\cite{Monteiro:2022nqt}, where a set of formally-defined currents related to the usual classical Lax pairs were shown to be not conserved in the presence of the quantum corrections.

One feature that may appear puzzling is that the anomaly \eqref{eq:infanom} of the currents, to give the example of SDG and in particular say the spin-1 current $j_\mu$, is not of the form of the chiral anomalies well-known from fermions on non-trivial backgrounds, briefly reviewed in section~\ref{sec:reviewchiral}. Naively, one may think that this is due to the fact that we are dealing with chiral transformations of gauge bosons, instead of those of fermions. However, the U(1) transformations we consider are precisely of the type that underlies the famous electric-magnetic duality of the photon field in Maxwell theory, as well as in Born-Infeld theory and other duality-obeying non-linear theories of electromagnetism. It was shown in \cite{Agullo:2016lkj} that the electric-magnetic duality is anomalous on a curved spacetime background, with the anomaly taking the expected form equivalent to \eqref{eq:fermanomgrav} up to a numerical factor. So the reason for the different type of chiral anomaly is not that we are dealing with chiral transformations of gauge bosons instead of fermions, but that the setting is simply distinct, including the precise manner in which the anomaly arises from the path integral. We are considering self-interacting gauge bosons on a trivial background. There is in fact an analogue in the case of photons. In Born-Infeld theory, where the photons are self-interacting, the electric-magnetic duality is broken at one loop on a trivial background. The anomalous amplitudes take a simple form somewhat reminiscent of those in SDYM and in SDG, though the details of the interactions and the amplitudes differ \cite{Elvang:2019twd}.

\subsection{Fermionic currents: 4D versus 2D}
\label{eq:fermion4D2D}

Here, we will take the approach of the usual chiral anomaly of fermions, and will ask the question of whether there is any special property of the symmetry currents of fermions on a self-dual gauge field or gravity background. Generally speaking, the 4D self-dual theories present some similarities to 2D theories, as is well known regarding integrability, so one may expect that such similarities can also be found in the context of fermions on self-dual backgrounds.

In 2D, quantum electrodynamics with a massless fermion, often called the Schwinger model, yields the following exact result \cite{Jackiw:1983nv} (page 275):
\begin{align}
\label{eq:fermPI2D}
Z[A]\, &=\int \mathscr{D}\psi \mathscr{D}\bar\psi \, \exp
\left( \int d^2x\,
\bar\psi (i \slashed{\partial}+\slashed{A}) \psi
\right) \nonumber \\
&= \exp\left(\frac1{2\pi}\int d^2x\, (\epsilon^{\mu\nu}F_{\mu\nu}) \frac1{\square_0} (\epsilon^{\mu\nu}F_{\mu\nu}) \right) \,.
\end{align}
This means that the vacuum expectation value of the U(1) vector current $\bar\psi \gamma_\mu \psi$ is
\[
\label{eq:jV2D}
j_\mu^\text{V} = \frac{\delta}{\delta A^\mu} \log Z[A] = \frac1{2\pi} \epsilon_{\mu\nu}\partial^\nu \frac1{\square_0} (\epsilon^{\alpha\beta}F_{\alpha\beta})\,.
\]
Clearly, this non-local current is conserved. Now, a crucial feature of 2D is that the chiral (or axial) current $i\bar\psi \gamma_\mu \gamma_5 \psi$ is dual to the vector current. This follows from the fact that $\epsilon^{\mu\nu}\gamma_\nu=i\gamma^\mu\gamma_5$. The resulting relation
\[
\label{eq:jVjA2D}
j_\mu^\text{V} = \epsilon_{\mu\nu}\, j^{\text{A},\nu}\,, \qquad\quad j_\mu^\text{A} = \frac1{2\pi}\, \partial_\mu \frac1{\square_0}(\epsilon^{\alpha\beta}F_{\alpha\beta})\,,
\]
leads to the expected 2D chiral anomaly,\, $\partial^\mu j_\mu^\text{A}= \frac1{2\pi} (\epsilon^{\alpha\beta}F_{\alpha\beta})$\,. We are considering here the U(1) gauge group. For SU($N$), it turns out that $Z[A]=1$, and therefore the currents have vanishing vacuum expectation value, because $\epsilon^{\mu\nu}F_{\mu\nu}$ is substituted by $\text{tr}(\epsilon^{\mu\nu}F_{\mu\nu})=0$.

Nevertheless, the 2D currents above have close analogues in 4D for self-dual gauge fields with the restricted set of gauge groups discussed in previous sections. For these, we have
\begin{align}
\label{eq:fermPI4D}
Z[A]\, &=\int \mathscr{D}\psi \mathscr{D}\bar\psi \, \exp
\left( \int d^4x\,
\bar\psi_\text{i} (i \slashed{\partial}+\slashed{A})_\text{ij} \psi_\text{j}
\right) \nonumber \\
&= \exp\left(\frac{a_\mathfrak{g}^2}{8} \int d^4x\; \Tr(F_{\mu\nu}F^{\mu\nu})
\,\frac1{\square_0^2}\,\Tr(F_{\mu\nu}F^{\mu\nu}) \right)\nonumber \\
&= \exp\left(-\frac{a_\mathfrak{g}^2}{32} \int d^4x\; \Tr(\varepsilon^{\mu\nu\rho\lambda}F_{\mu\nu}F_{\rho\lambda})
\,\frac1{\square_0^2}\,\Tr(\varepsilon^{\mu\nu\rho\lambda}F_{\mu\nu}F_{\rho\lambda}) \right)\,.
\end{align}
The second equality follows from \eqref{eq:qcSSDYM} and the discussion above that equation, particularly the fact that the effect of the fermionic loop is minus that of the gluonic loop (or the complex scalar loop). The final equality follows from \eqref{eq:SDFF} for self-dual fields. The result is clearly analogous to the 2D story, and this extends to the currents. We have, for the vector current $\bar\psi_\text{i} \gamma_\mu T^a_\text{ij}\psi_\text{j}$, the vacuum expectation value
\begin{align}
j_\mu^{\text{V},a} & = \frac{\delta}{\delta A^{a,\mu}} \log Z[A] = 
\frac{a_\mathfrak{g}^2}{8}\;
\varepsilon_{\nu\mu\rho\lambda}\left( (D^\nu F^{a,\rho\lambda})+F^{a,\rho\lambda}\partial^\nu
\right)\frac1{\square_0^2}\,\Tr(\varepsilon^{\alpha\beta\gamma\delta}F_{\alpha\beta}F_{\gamma\delta}) \nonumber \\
& = -\frac{a_\mathfrak{g}^2}{8}\;
\varepsilon_{\mu\nu\rho\lambda} F^{a,\rho\lambda}\,\partial^\nu\,
\frac1{\square_0^2}\,\Tr(\varepsilon^{\alpha\beta\gamma\delta}F_{\alpha\beta}F_{\gamma\delta})
\,,
\end{align}
where the second line follows from the Bianchi identity.
This current is analogous to the 2D current \eqref{eq:jV2D}.
Its conservation is also easily checked, using again the Bianchi identity:
\[
D^\mu j_\mu^{\text{V},a}=-\frac{a_\mathfrak{g}^2}{8}\;
\varepsilon_{\mu\nu\rho\lambda} \left((D^\mu F^{a,\rho\lambda})\partial^\nu+F^{a,\rho\lambda}\partial^\mu\partial^\nu\right)
\frac1{\square_0^2}\,\Tr(\varepsilon^{\alpha\beta\gamma\delta}F_{\alpha\beta}F_{\gamma\delta})
=0\,.
\]
The 2D story suggests that we may be able to identify the vacuum expectation value of the singlet axial current $i\bar\psi_\text{i} \gamma_\mu \gamma_5 \psi_\text{i}$ such that we have a relation analogous to \eqref{eq:jVjA2D}:\footnote{Here, taking $G(x,y)$ to be the Green's function defining $1/\square$, we assumed that the following integration by parts is allowed:
\begin{align}
    \partial_\nu \frac1{\square} f(x) &= \int d^4y\, f(y)\frac{\partial}{\partial x^\nu} G(x,y)  = \int d^4y\,f(y) \frac1{\square_x} \frac{\partial}{\partial x^\nu} \,\delta(x,y)  = -\int d^4y\,f(y) \frac1{\square_x} \frac{\partial}{\partial y^\nu} \,\delta(x,y) \nonumber \\ & = \int d^4y\,\frac1{\square_x} \,\delta(x,y) \frac{\partial}{\partial y^\nu}f(y) = \int d^4y\,G(x,y) \frac{\partial}{\partial y^\nu}f(y) = \frac1{\square}\partial_\nu f(x)\,. \nonumber 
    \end{align}}
\[
\label{eq:jVjA4D}
j_\mu^{\text{V},a} = -2\pi^2a_\mathfrak{g}^2\;
\varepsilon_{\mu\nu\rho\lambda} \left(F^{a,\rho\lambda}\,
\frac1{\square_0}\right)\,j^{\text{A},\nu}\,, \qquad j^{\text{A}}_\mu = \frac1{16\pi^2} \,\partial_\mu \,\frac1{\square_0} \Tr(\varepsilon^{\alpha\beta\gamma\delta}F_{\alpha\beta}F_{\gamma\delta})\,.
\]
This is consistent with the 4D chiral anomaly,
\[\partial^\mu j^{\text{A}}_\mu=\frac1{16\pi^2}\Tr(\varepsilon^{\alpha\beta\gamma\delta}F_{\alpha\beta}F_{\gamma\delta}).
\]
In typical treatments of the chiral anomaly, the axial current is not computed --- what is computed is its gauge-invariant anomaly. We propose that in the setting of a self-dual background, a gauge-invariant regularised singlet axial current is possible, given as above in analogy to 2D.

It would be interesting to explore further this 4D/2D analogy to see if we can import to 4D (even if only in the self-dual sector) the wealth of exact results known for the Schwinger model. The gravitational analogue merits attention too, though we leave this for future consideration.


\section{Insights from the connection to the trace anomaly}
\label{sec:insights}


\subsection{Properties of the Fradkin-Tseytlin-Paneitz operator}
\label{sec:Paneitz}

The trace anomaly gives us clues into the appearance of a fourth-order (inverse) differential operator in the quantum-corrected actions for SDYM and SDG. The FTP operator $\Delta_4$ defined in \eqref{eq:Paneitz} is such that $\sqrt{|g|}\Delta_4$ is Weyl invariant, and satisfies $\Delta_4=\square^2$ on Ricci-flat backgrounds. This operator exposes the conformal properties of the theories.

One important property of quantum SDYM is the conformal invariance of its loop amplitudes, given in \eqref{eq:An}. We recall that these are the all-plus one-loop amplitudes of full Yang-Mills theory. The conformal invariance was observed empirically in \cite{Henn:2019mvc}, by noting that the generators of the conformal algebra --- conveniently expressed in spinor-helicity notation \cite{Witten:2003nn} --- eliminate the amplitudes. A brief review of this story is presented in Appendix~\ref{appendixSDYM}.\footnote{Ref.~\cite{Henn:2019mvc}'s full proof of the conformal invariance of the amplitudes, not included in the appendix, noted that the amplitudes can be written as a sum of simple terms that are separately conformally invariant. Interestingly, these same terms had appeared as BCJ numerators of one-loop MHV amplitudes in ${\mathcal N}=4$ super-Yang-Mills theory \cite{He:2015wgf}. Indeed, a connection between the maximally supersymmetric theory and the self-dual sector of the non-supersymmetric theory has long been known \cite{Bern:1996ja}; see \cite{Britto:2020crg} for a recent proof of this relationship. This in turn suggests that the structures studied in our paper play some role in ${\mathcal N}=4$ super-Yang-Mills theory.}

The form of the quantum-corrected action for SDYM in \eqref{eq:qcSSDYM} provides an explanation for the conformal invariance of these amplitudes --- though only for the gauge groups where this action is valid, e.g.~SU(3). The relation of that action to the twistor-space theory of \cite{Costello:2021bah} already indicates the presence of conformal symmetry, but let us focus here on the spacetime formulation. Consider the following action on a general background:
\[
\label{eq:SDYMWeyl}
S_{\text{SDYM}} -\frac{a_\mathfrak{g}^2}{8} \int d^4x\,\sqrt{|g|}\; \Tr(F_{\mu\nu}F^{\mu\nu})
\,\frac1{\Delta_4}\,\Tr(F_{\mu\nu}F^{\mu\nu})\,.
\]
This action is clearly invariant under a Weyl transformation,\footnote{Notice that $\chi=\frac1{\Delta_4}\,\Tr(F_{\mu\nu}F^{\mu\nu})$ is Weyl invariant, as implied by $\Delta_4\, \chi = \Tr(F_{\mu\nu}F^{\mu\nu})$.}
\[
\label{eq:WeylF}
g_{\mu\nu} \to e^{2\Omega}g_{\mu\nu}\,,\quad
F_{\mu\nu}F^{\mu\nu} \to e^{-4\Omega}F_{\mu\nu}F^{\mu\nu}\,,\quad
\Delta_4 \to e^{-4\Omega}\Delta_4\,.
\]
Moreover, it reduces to \eqref{eq:qcSSDYM} in the flat background case. This straightforwardly explains the conformal invariance of the quantum-corrected theory on flat spacetime, and hence of the scattering amplitudes. To see this, we recall that a conformal transformation $x^\mu\to x'^\mu(x)$ is such that
\[
\label{eq:conformal}
\eta_{\mu\nu}dx'^\mu dx'^\nu=e^{2\Omega(x)}\,\eta_{\mu\nu}dx^\mu dx^\nu\,.
\]
Here, unlike in \eqref{eq:WeylF}, $\Omega$ is a restricted function, defined by the parameters of 4D conformal transformations. Now, Weyl invariance implies that the flat-spacetime action takes the same form \eqref{eq:qcSSDYM} whether the coordinate choice is $x^\mu$ or $x'^\mu$; that is, the factor $e^{2\Omega(x)}$ in \eqref{eq:conformal} cancels out in the action. This is the statement of conformal invariance of a theory living on flat spacetime. In fact, one does not need to appeal to the full Weyl-invariance of the action \eqref{eq:SDYMWeyl} on a curved background. The action \eqref{eq:qcSSDYM} is already invariant under conformal transformations. Since $\Delta_4=\Box_0^2$ for flat spacetime, $\Box_0^2$ has the restricted Weyl covariance associated to the allowed choices of $\Omega$ in \eqref{eq:conformal}; that is, it is conformally covariant. Explicitly, acting on a scalar $\phi'(x')=\phi(x)$, we have under the conformal transformation:
\[
(\partial_{x'}^2)^2 = e^{-4\Omega} (\partial_{x}^2)^2\,,
\]
where we denote\, $\partial_{x'}^2=\frac{\partial}{\partial x'^\mu}\frac{\partial}{\partial x'_\mu}$\,, etc; the restriction on $\Omega$ allows the conformal factor to `factor out' on the right-hand side. 

While the operator $\Delta_4$ is not strictly required to explain the conformal invariance of the amplitudes, as we have just explained, it is important when considering the extension to curved backgrounds, to be discussed below. Furthermore, as mentioned previously, the complete SDYM quantum-corrected action is not known explicitly for SU($N$), despite partial progress in \cite{Monteiro:2022nqt}. It is likely that conformal invariance will be a useful guide to obtaining the complete quantum correction.

It is worthwhile to mention that, in the gravity case \eqref{eq:qcSSDG}, we have
\[
\label{eq:SDGsquaretoDelta}
\int d^4x\,\sqrt{|g|}\; C^2 \,\frac1{\square^2}\,C^2 \;=\;
\int d^4x\,\sqrt{|g|}\; C^2 \,\frac1{\Delta_4}\,C^2
\]
under the self-duality constraint imposed by $S_{\text{SDG}}$, which implies Ricci flatness. Indeed, as we describe in the Appendix~\ref{appendixSDG}, the four-point gravity amplitude \eqref{eq:M4} is almost conformal invariant (almost because of the dimensionful coupling) due to the Weyl invariance of the right-hand side of \eqref{eq:SDGsquaretoDelta}. The higher-point amplitudes do not share this property, as a result of $S_{\text{SDG}}$ not being conformally invariant.


\subsection{Quantum-corrected actions on non-trivial backgrounds}
\label{sec:backgrounds}

The study of field theory on curved backgrounds has received a new push in recent years, inspired by novel methods, not least those originating in the amplitudes literature. In the previous section, we mentioned already the possibility of extending the non-local quantum-corrected action for SDYM to non-trivial background spacetimes, in the form
\[
\label{eq:backqcSSDYM}
S_{\text{q.c.SDYM}} = S_{\text{SDYM}} -\frac{a_\mathfrak{g}^2}{8} \int d^4x\,\sqrt{|g|}\; \Tr(F_{\mu\nu}F^{\mu\nu})
\,\frac1{\Delta_4}\,\Tr(F_{\mu\nu}F^{\mu\nu}) + \ldots\,.
\]
This a natural conjecture for such an extension, because (i) Weyl invariance is a natural upgrade for the conformal invariance of quantum SDYM exhibited by its flat-spacetime scattering amplitudes, and (ii) the FTP operator $\Delta_4$ makes an appearance already in the quantum correction associated to the trace anomaly, also when considering a curved background. The action above is subject to the restriction on the gauge group mentioned earlier, e.g. SU(3). The ellipsis accounts for possible terms that have no counterpart in the flat spacetime limit. Examples where this conjecture could be tested are backgrounds with significant symmetry, such as plane waves and (anti-)de Sitter; see e.g.~\cite{Adamo:2017nia,Adamo:2017sze,Adamo:2018mpq,Adamo:2019zmk,Adamo:2020qru,Armstrong:2020woi,Gomez:2021ujt,Armstrong:2022mfr,Lipstein:2023pih} for recent work along these lines. Another natural setting is provided by SDG backgrounds \cite{Adamo:2020syc,Adamo:2021bej,Bittleston:2023bzp,Adamo:2023zeh,Adamo:2023fbj}.

In the case of SDG, one may ask how \eqref{eq:qcSSDG} would extend to SDG-$\Lambda$, i.e.~allowing for a cosmological constant. In SDG-$\Lambda$, the self-duality condition applies to the Weyl tensor only, while the Ricci tensor satisfies the Einstein-$\Lambda$ equations, $R_{\mu\nu}=\Lambda g_{\mu\nu}$. See \cite{Lipstein:2023pih} for a recent discussion of the second Plebanski equation with non-zero $\Lambda$. The $\Lambda$-self-duality condition is imposed by the `tree-level part' of the action, $S_{\text{SDG-}\Lambda}$. For the quantum correction, a natural candidate is suggested by \eqref{eq:SDGsquaretoDelta}, namely
\[
\label{eq:backqcSSDG}
S_{\text{q.c.SDG-}\Lambda} = S_{\text{SDG-}\Lambda} -\frac{a_\text{G}^2}{8} \int d^4x\,\sqrt{|g|}\; C^2 \,\frac1{\Delta_4}\,C^2 + \ldots\,.
\]
Recalling the definition of the FTP operator in \eqref{eq:Paneitz}, under $R_{\mu\nu}=\Lambda g_{\mu\nu}$ we have
\[
\Delta_4=\square^2-\frac{2}{3}\Lambda\square. 
\]
The light-cone version of $S_{\text{SDG-}\Lambda}$ obtained in \cite{Lipstein:2023pih} provides a starting point for testing this proposal, by performing explicitly a loop calculation, though this is not straightforward on the non-trivial background.

Another question is how this story extends to self-dual conformal gravity (see e.g.~\cite{Dunajski:2013zta}). A natural expectation is that the quantum correction in \eqref{eq:backqcSSDG} is again present. Yet another class of related theories are the chiral higher-spin theories of \cite{Metsaev:1991mt,Metsaev:1991nb,Ponomarev:2016lrm,Ponomarev:2017nrr}, which generalise SDG and SDYM to include massless fields of higher spins. For some theories in this class, there exist non-vanishing one-loop amplitudes \cite{Skvortsov:2018jea,Skvortsov:2020wtf,Skvortsov:2020gpn}, and the same quantum correction occurs, involving only amplitudes with external gravitons/gluons (no higher spins); the coefficient of the correction is then easily adjusted \cite{Adamo:2022lah,Monteiro:2022xwq}. Ref.~\cite{Monteiro:2022xwq} includes a summary of the status of loop amplitudes in chiral higher-spin theories.


\section{Relation to UV divergence in two-loop pure gravity}
\label{sec:UVgrav}

In this section, we will move beyond the self-dual sectors to comment on an intriguing feature of the ultraviolet behaviour of pure gravity. The well-known two-loop divergence \cite{Goroff:1985sz,Goroff:1985th,vandeVen:1991gw} has been investigated more recently using modern unitarity methods \cite{Bern:2015xsa,Bern:2017puu,Abreu:2020lyk}. It was argued in \cite{Bern:2015xsa} that the coefficient of the dimensionally-regularised divergence $\frac1{D-4}$, which can be cancelled with a (Riemann${}^3$) counterterm, is not directly physical; the reason being that it depends on a possible coupling to 3-form fields, which should have no physical consequence as they do not propagate in strictly four spacetime dimensions. Instead, the physical effect of the divergence is the renormalisation-scale dependence, signalled by a $\log \mu^2$ term in the two-loop amplitude, which in the case of gravity is not tied to the $\frac1{D-4}$ ultraviolet divergence. 

\begin{figure}[t]
\centering
\includegraphics[width=0.5\textwidth]{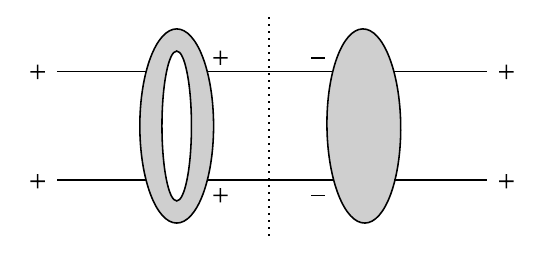}
\caption{Example of a two-particle cut which enters the computation of the $\log \mu^2$ term in the two-loop all-plus pure gravity amplitude at four-points. The internal legs crossing the dotted line are on-shell.}
\label{fig:cut}
\end{figure}

Noticing this fact, refs.~\cite{Bern:2015xsa,Bern:2017puu} computed the kinematic coefficient of the $\log \mu^2$ term via unitarity cuts, for the four-point two-loop amplitude. They focused on the all-plus helicity case, which is easier to tackle with the unitarity method due to the fact that only two-particle cuts contribute to $\log \mu^2$. To obtain these cuts, one glues a four-point all-plus one-loop amplitude to a four-point tree-level (MHV) amplitude, as in Figure~\ref{fig:cut}. From the reasoning of the cuts, it is shown in \cite{Bern:2015xsa,Bern:2017puu} that the coefficient of $\log \mu^2$ does not depend on the number of 3-form fields. In addition, they observe that, beyond pure gravity (i.e.~with any number of bosons and fermions in the loop of the one-loop side of the cut), the coefficient is proportional to $N_\text{bosons}-N_\text{fermions}$. This follows directly from the properties of all-plus one-loop amplitudes mentioned in earlier sections; in this setting, such a four-point amplitude appears on one side of the cut. An obvious consequence is that the $\log \mu^2$ term is absent for supergravity theories; indeed, supersymmetry guarantees ultraviolet finiteness at two loops \cite{Grisaru:1976nn}.

In section~\ref{sec:sdanomaliestr}, we discussed the twistor-derived bosonic models of \cite{Costello:2021bah,Bittleston:2022nfr} for `improved' SDYM and SDG. These models achieve the cancellation of the one-loop amplitudes of ordinary SDYM and SDG, in a very different manner from the cancellation achieved in the supersymmetric extensions of SDYM and SDG (where the one-loop amplitudes are also absent). A natural question, given the relevance of the all-plus one-loop amplitudes to the ultraviolet behaviour in pure gravity, is whether there is an analogous bosonic mechanism that ultimately cancels the two-loop divergence, as an alternative to supersymmetry. Taking the perspective of the non-local action (but one could also consider the local form with an auxiliary scalar), our observation is that the action
\[
\label{eq:EHfix}
S_{\text{EH}} + \frac{a_\text{G}^2}{8} \int d^4x\,\sqrt{|g|}\; C^2 \,\frac1{\Delta_4}\,C^2 +\;\ldots
\]
for pure gravity provides such a cancellation of the $\log \mu^2$ term. It leads to the vanishing (by cancellation between the Einstein-Hilbert term and the non-local term) of the one-loop all-plus amplitudes. This is just the cancellation discussed in section~\ref{sec:sdanomaliestr}. Recalling Figure~\ref{fig:cut}, this eliminates the four-point amplitude in the one-loop side of the cut, and following the argument of \cite{Bern:2015xsa,Bern:2017puu}, it leads to a vanishing coefficient for $\log \mu^2$ in the two-loop all-plus four-point amplitude. That is, the one-loop non-local term is fixing the two-loop behaviour in the case of all-plus amplitudes. 

The ellipsis (\ldots) above indicates the fact that the non-local term written here is certainly not sufficient for eliminating the $\log \mu^2$ term in all two-loop four-point amplitudes; while it is sufficient for the all-plus two-loop case, the one-minus case also requires attention (the two-plus/two-minus case has no $\log \mu^2$ term \cite{Abreu:2020lyk}). The non-local term we have proposed in \eqref{eq:EHfix} cannot do the job for the one-minus two-loop amplitude, because it clearly vanishes when evaluated on the one-minus one-loop amplitude\footnote{We can see this as follows. At linearised level for $\pm$ helicity and on-shell momentum $k$, we have $C^\pm_{\mu\nu\rho\lambda}\sim \varepsilon^\pm_{[\mu} k_{\nu]}\varepsilon^\pm_{[\rho} k_{\lambda]}$. Hence, $C_{\mu\nu\rho\lambda}(1^-)\,C^{\mu\nu\rho\lambda}(2^+)\sim (\varepsilon^-_1{}_{[\mu} k_1{}_{\nu]}\varepsilon^+_2{}^{[\mu} k_2{}^{\nu]})^2$, which vanishes because $\varepsilon^+_2{}^{[\mu} k_2{}^{\nu]}$ and $\varepsilon^-_1{}^{[\mu} k_1{}^{\nu]}$ are self-dual and anti-self-dual tensors, respectively.} that would appear in the 2-particle cuts. In addition, for the $\log \mu^2$ term of the one-minus two-loop amplitude, there are also 3-particle cuts, involving five-point tree amplitudes on both sides of the cut. We refer the reader to \cite{Bern:2017puu} for details. One would also have to consider whether a relation between the two-loop $\log \mu^2$ term and the one-loop all-plus and one-minus amplitudes persists at higher points, and of course at higher loops.

While this could be considered a long shot, we see no obstacle of principle that prevents us from following this procedure for at least the four-point two-loop gravity amplitudes, though we have not determined the precise form of the required terms for the ellipsis in \eqref{eq:EHfix}. The physical origin of why such a non-local term should be present in the action remains unclear. We merely note its potential consequence.


\section{Conclusion}
\label{sec:conclusion}

The main goal of this paper was to connect the non-local quantum-corrected actions of self-dual Yang-Mills theory and gravity, to the non-local actions that have long appeared in the context of anomalies. With the benefit of hindsight, the quantum corrections to SDYM and SDG \cite{Costello:2021bah,Bittleston:2022nfr,Monteiro:2022nqt} could have been determined by simple analogy with the trace anomaly: by trying a natural non-local functional, and then fixing the coefficient by matching to an amplitude. In addition, we discussed how the quantum corrections in the self-dual theories are related to chiral anomalies. On this point, we have provided two distinct points of view. The first is that the quantum correction signals the anomaly of the chiral U(1) symmetry acting on the gluon or the graviton in the self-dual theories; this is a symmetry associated to the `electric-magnetic' duality of the linearised fields. The second is that the quantum correction can be interpreted from the integrating out of a fermion. We noted that there is a close analogy with the long-known results of the 2D Schwinger model, which merits further investigation.

One clue from the trace anomaly literature is the appearance of the Fradkin-Tseytlin-Paneitz fourth-order differential operator $\Delta_4$, which allowed us to (i) understand in a simple manner the conformal properties of the loop-level scattering amplitudes in the self-dual theories, and (ii) present natural conjectures for extending the quantum-corrected actions to non-trivial spacetime backgrounds. This aspect of our work is part of ongoing efforts to bring modern amplitudes-related constructions to the study of quantum field theories on curved backgrounds.

Regarding Yang-Mills theory, we proposed a Weyl-invariant formulation of the quantum-corrected self-dual sector on curved spacetimes. This is, of course, a part of the full quantum-corrected Yang-Mills action. It would be interesting to explore whether the rational parts of loop amplitudes beyond the self-dual sector can be written in terms of analogous non-local quantum corrections. Such rational terms can be efficiently computed using unitarity methods, at least at one-loop \cite{Badger:2008cm}. We also remarked on the relationship between the anomaly of the self-dual sector in pure Yang-Mills theory, and the one-loop MHV sector of ${\mathcal N}=4$ super-Yang-Mills theory, pointed out long ago \cite{Bern:1996ja}. It would be nice to understand the full extent of this relationship, e.g. whether it extends to non-trivial backgrounds.

We briefly discussed how the anomaly of the self-dual sector in gravity suggests a one-loop effective mechanism for altering the pathological ultraviolet behaviour of pure gravity, at least at two loops --- a mechanism that is an alternative to supersymmetry. This suggestion relies on an apparent connection between rational one-loop amplitudes and the divergent behaviour of two-loop amplitudes. Our observation builds on modern studies of the two-loop divergence in \cite{Bern:2015xsa,Bern:2017puu,Abreu:2020lyk}, and also on what appears to be a close connection between anomalies and ultraviolet divergences in pure (super)gravity more generally \cite{Carrasco:2013ypa,Bern:2013uka,Bern:2017rjw,Bern:2019isl,Carrasco:2022lbm}. It would be important to understand better this intriguing connection.

Partly based on this observation, and partly based on the role of a fourth-order differential operator, it is natural to ask whether there is an unexplored connection to the vast topic of higher-derivative gravity. Quadratic gravity, recently reviewed in \cite{Donoghue:2021cza,Salvio:2018crh}, has been proposed as a renormalisable theory of quantum gravity, despite the fact that it seems at odds with common expectations about quantum field theory, regarding e.g.~causality and the appearance of $1/p^4$ propagators, in apparent violation of unitarity. Such propagators arise explicitly from the non-local quantum corrections we have discussed here. Nevertheless, the loop amplitudes in the self-dual sector turn out to have only simple poles in the Mandelstam variables, instead of the quadratic poles naively expected from $1/p^4$, due to kinematic identities. While the self-dual theory is non-unitary on its own (as it corresponds to a complex truncation of the Hamiltonian of the full theory), it is also a sector of a unitary theory --- certainly in the Yang-Mills case. The quantum correction in self-dual gravity, with its $1/p^4$ propagator, should also be a part of the quantum corrections in general relativity, seen as an effective theory. This suggests that such propagators may not always represent a terminal pathology. Finally, it is worth mentioning that conformal gravity has been argued to be a limit of quadratic gravity \cite{Salvio:2017qkx}, and has an interesting double copy construction~\cite{Johansson:2017srf,Johansson:2018ues}. It is tempting to ask whether there is a common ground between these topics and the aforementioned relation of the ultraviolet behaviour of pure gravity to the anomaly of the self-dual sector, the latter being reminiscent of the trace anomaly.

\section*{Acknowledgements}

We thank Tim Adamo, Chandramouli Chowdhury, Anton Ilderton, Henrik Johansson, Arthur Lipstein, Lionel Mason, Silvia Nagy and Kajal Singh for related discussions. RM is supported by the Royal Society via a University Research Fellowship. GD is supported by the Royal Society via a studentship grant. The work of SW is funded by the Knut and Alice Wallenberg Foundation grant KAW 2021.0170 and the Olle Engkvists Stiftelse grant 2180108. We also received support from the Science and Technology Facilities Council (STFC) Consolidated Grants ST/T000686/1 “Amplitudes, Strings \& Duality” and ST/X00063X/1 “Amplitudes, Strings \& Duality”. No new data were generated or analysed during this study.

\appendix
\section{Conformal properties of the scattering amplitudes}
\label{appendix}

In this appendix, we first recall the action of the generators of the conformal algebra on scattering amplitudes; then review the result in \cite{Henn:2019mvc} regarding the conformal invariance of the all-plus one-loop amplitudes in Yang-Mills theory. We also address the analogous question in gravity.

The generators of the four-dimensional conformal group for an $n$-particle amplitude, expressed in spinor-helicity language, are as follows \cite{Witten:2003nn}: the Lorentz generators,  
\begin{equation}
    \begin{aligned}
    J_{\alpha \beta} &= \frac{i}{2} \sum_{i=1}^n \left(\lambda_{i\alpha} \frac{\partial}{\partial \lambda^{\beta}_i} +\lambda_{i\beta} \frac{\partial}{\partial \lambda^{\alpha}_i} \right), \\
    \tilde{J}_{\alphadot \betadot} &= \frac{i}{2} \sum_{i=1}^n \left(\lambdat_{i\alphadot} \frac{\partial}{\partial \lambdat^{\betadot}_i} +\lambdat_{i\betadot} \frac{\partial}{\partial \lambdat^{\alphadot}_i}\right),
\end{aligned}
\end{equation}
the momentum generator,
\begin{equation}
    P^{\alpha \alphadot} = \sum_{i=1}^n \lambda^\alpha_i \lambdat^{\alphadot}_i,
\end{equation}
the dilatation generator,
\begin{equation}
    D= i \sum_{i=1}^n\left( \frac{1}{2}\lambda^\alpha_i \frac{\partial}{\partial \lambda^{\alpha}_i} +\frac{1}{2}\lambdat^{\alphadot}_i \frac{\partial}{\partial \lambdat^{\alphadot}_i} +1 \right), 
\end{equation}
and the special conformal generator, 
\begin{equation}
    K_{\alpha \alphadot}=\sum_{i=1}^n \frac{\partial^2}{\partial \lambda^\alpha _i \partial \lambdat^{\alphadot}_i}.
\end{equation}
Consider an amplitude of the form
\begin{equation}
    \mathcal{A}(\{p_i \}) = \delta^4 \left(\sum_{i=1}^n p_i \right) A(\{p_i \}).
\end{equation}
The Poincar\'{e} generators act on this expression as
\begin{equation}
    \begin{aligned}
    J_{\alpha \beta} \mathcal{A} &= \delta^4 \left(P \right) J_{\alpha \beta} A , \\
    \tilde{J}_{\alphadot \betadot}\mathcal{A} &= \delta^4 \left(P \right) \tilde{J}_{\alpha \beta} A, \\
    P^{\alpha \alphadot} \mathcal{A} &= 0.
\end{aligned}
\end{equation}
Thus, Poincar\'{e} invariance follows from the Lorentz invariance of the `delta-stripped' amplitude. This, in turn, follows from the fact the stripped amplitude can be written as a function of spinor brackets $\lambda_{i\alpha}\lambda_j{}^\alpha$ and $\lambdat_{i\alphadot}\lambdat_j{}^{\alphadot}$. The dilatation operator acts on the amplitude as 
\begin{equation}
    D \mathcal{A} = \delta^4(P) (-4iA+DA),
\end{equation}
where the first term arises from differentiating the delta function as a distribution. Dilatation invariance requires $DA=4iA$, which holds for a dimensionless coupling. This is because $-iDA=(d_A-d_c+n)A$, where $d_A$ is the mass dimension of $A$, which is $4-n$ in four dimensions, $d_c$ is the mass dimension of the coupling constant factor in $A$, and $n$ arises from the $+1$ in the summand defining the generator $D$; hence, $-iDA=(4-d_c)A$ and
\[
D \mathcal{A}=-i\,d_c\,\mathcal{A}.
\]
That is, dilatation invariance corresponds to a dimensionless coupling.
Imposing Lorentz invariance, $J_{\alpha \beta} A=0=\tilde{J}_{\alphadot \betadot} A$, the special conformal generator acts on the amplitude as 
\begin{align}
    K_{\alpha \alphadot} \mathcal{A} &= \delta^4(P)K_{\alpha \alphadot} A + \frac{\partial \delta^4(P)}{\partial P^{\alpha \alphadot}} (-4A-iDA) \nonumber \\
    &= \delta^4(P)K_{\alpha \alphadot} A - d_c \,\frac{\partial \delta^4(P)}{\partial P^{\alpha \alphadot}}\, A.
\end{align}
We see that special conformal invariance requires that $ K_{\alpha \alphadot} A =0$, in addition to dilatation invariance.

\subsection{All-plus one-loop Yang-Mills amplitudes}
\label{appendixSDYM}

Ref.~\cite{Henn:2019mvc} showed that the all-plus one-loop Yang-Mills amplitudes, first obtained in closed form in \cite{Bern:1993qk}, and which lie in the self-dual sector, exhibit conformal invariance. That is,
\[
 P^{\alpha \alphadot} \mathcal{A}^{(1)}_\text{YM} = 0,\;
 J_{\alpha \beta} \mathcal{A}^{(1)}_\text{YM} = 0,\;
 \tilde{J}_{\alphadot \betadot}\mathcal{A}^{(1)}_\text{YM} =0,\;
 D \mathcal{A}^{(1)}_\text{YM} = 0,\;
 K_{\alpha \alphadot} \mathcal{A}^{(1)}_\text{YM} = 0.
\]
Only the special conformal invariance is non-trivial. It is most easily seen in the very simple four-point case, where
\begin{equation}
\mathcal{A}^{(1)}_\text{YM}(1^{+},2^{+},3^{+},4^{+})= \delta^4(p_1+p_2+p_3+p_4) A^{(1)}_\text{YM}(1^{+},2^{+},3^{+},4^{+}),
\end{equation}
where 
\begin{equation}
    A^{(1)}_\text{YM}(1^{+},2^{+},3^{+},4^{+}) = -\frac{i\,g^4}{48 \pi^2}\; \frac{[12][34]}{\langle 12\rangle \langle 34\rangle }
\end{equation}
for the colour-ordered amplitude.

\subsection{All-plus one-loop gravity amplitudes}
\label{appendixSDG}

The all-plus one-loop gravity amplitudes, which lie in the self-dual sector, were first obtained in closed form in \cite{Bern:1998xc}. The four-point case is
\begin{equation}
\mathcal{A}^{(1)}_\text{G}(1^{+},2^{+},3^{+},4^{+})= \delta^4(p_1+p_2+p_3+p_4) A^{(1)}_\text{G}(1^{+},2^{+},3^{+},4^{+}),
\end{equation}
with
\begin{equation}
\label{eq:M4k}
    A^{(1)}_\text{G}(1^{+},2^{+},3^{+},4^{+}) = -\frac{i}{120 (4\pi)^2} \left(\frac{\kappa}{2} \right)^4 (s_{12}^2 +s_{23}^2 +s_{31}^2) \left(\frac{s_{12}s_{23}}{\langle 12 \rangle \langle 23 \rangle \langle 34 \rangle \langle 41 \rangle} \right)^2,
\end{equation}
where we made the coupling explicit.
The Poincar\'e invariance of the amplitudes is clear,
\[
 P^{\alpha \alphadot} \mathcal{A}_\text{G} = 0,\;
 J_{\alpha \beta} \mathcal{A}_\text{G} = 0,\;
 \tilde{J}_{\alphadot \betadot}\mathcal{A}_\text{G} =0.
\]
The $n$-point amplitude goes like $\kappa^n$, so that the dimensionful coupling breaks dilatation symmetry,
\[
D \mathcal{A}_\text{G}=i \, n \,\mathcal{A}_\text{G}\,.
\]
We then have also
\begin{equation}
    K_{\alpha \alphadot} \mathcal{A}_\text{G} = \delta^4(P)K_{\alpha \alphadot} A_\text{G} + n\,\frac{\partial \delta^4(P)}{\partial P^{\alpha \alphadot}} \,A_\text{G}.
\end{equation}

It turns out that the four-point amplitude \eqref{eq:M4k} is special because
\[
K_{\alpha \alphadot}\, A^{(1)}_\text{G}(1^{+},2^{+},3^{+},4^{+}) =0,
\]
a property that is not shared by higher-point amplitudes. Therefore, it is almost conformal invariant, with full invariance failing due to the dimensionful coupling. In terms of the discussion in the main body of the paper, we interpret this as being due to the following fact: in the action \eqref{eq:qcSSDG}, while the part $S_\text{SDG}$ is not conformally invariant, the quantum correction is; in fact, the latter can be written in a Weyl-invariant manner as in \eqref{eq:SDGsquaretoDelta}. The quantum correction provides the one-loop effective vertices, whose explicit form in light-cone gauge we wrote in \eqref{eq:SDGLC}, taken from \cite{Monteiro:2022nqt}. It is helpful to think of the Feynman rules resulting from the light-cone action. The four-point amplitude uses only the four-point one-loop effective vertex, with legs put on-shell, and therefore it is not surprising that there is a remnant of conformal symmetry in this particular amplitude. The five-point amplitude uses both the five-point one-loop effective vertex and, in a different set of diagrams, the four-point one-loop effective vertex with a three-point `tree vertex' attached, with the latter coming from the $S_\text{SDG}$ part of the action. We interpret the failure of the almost-conformal invariance for $n>4$ (that is, $K_{\alpha \alphadot}\, A^{(1)}_\text{G}\neq0$) as due to the appearance of the three-point vertex from the conformally-non-invariant piece $S_\text{SDG}$.

\bibliography{refs}
\bibliographystyle{JHEP}

\end{document}